\documentstyle[preprint,aps]{revtex}
\begin{document}
\draft
\title{Locally Anisotropic Gravity and Strings}
\author{Sergiu I. Vacaru}
\address{Institute of Applied Physics, Academy of Sciences of Moldova \\
5 Academy str., Chi\c{s}in\u{a}u 277028, Republic of Moldova}
\date{\today}
\maketitle
\begin{abstract}
 An introduction into the theory of locally anisotropic spaces (modelled as
 vector bundles provided with compatible nonlinear and distinguished linear
 connections and metric structures and containing  as particular cases
 different types of Kaluza--Klein and/or extensions of Lagrange and Finsler
 spaces ) is presented. The conditions for consistent propagation of closed
 strings in locally anisotropic background spaces are analyzed. The connection
 between the conformal invariance, the vanishing of the renormalization group
$\beta $--function of the generalized $\sigma$--model and field equations of
 locally anisotropic gravity is studied in detail.

{\bf \copyright \ S.I. Vacaru}
\end{abstract}
\vskip 15pt

\section{Introduction}

The relationship between two dimensional $\sigma $-models and strings has
been considered by several authors [1-5] in order to discuss the effective
low energy field equations for the massless models of strings. In this paper
we shall investigate some of the problems associated with the theory of
locally anisotropic strings being a natural generalization to locally
anisotropic (la) backgrounds (we shall write in brief la-backgrounds,
la-spaces and la-geometry) of the Polyakov's covariant functional-integral
approach to string theory [6]. Our aim is to show that a corresponding
low-energy string dynamics contains the motion equations for field equations
on la-spaces.

The first geometric models of la-spaces were formulated by P. Finsler [7],
studied in detail and generalized by A.Berwald [8] and E. Cartan [9]. The
geometry of Lagrange and Finsler spaces, its extensions and possible
applications in physics (locally anisotropic gauge and gravitational
theories, statistical physics and relativistic optics in locally anisotropic
media and other topics) were considered in a number of works, see references
from [10,11]. It seems likely that classical and quantum physical models
admit a straightforward extension of la-spaces with compatible metric and
connections structures. In this case we can define locally anisotropic
spinors [12], formulate the theory of locally anisotropic interactions of
gauge and gravitational fields [13] and consider superspaces with local
anisotropy [14]. Here we remark that modern Kaluza-Klein theories can be
considered as physical models on la-spaces with trivial nonlinear
connection stucture.

There are a number of arguments for taking into account effects of possible
local anisotropy of fundamental interactions. It is well known the result
that a self consistent description of radiational and dissipation processes
in classical and quantum field theories requires adding of higher derivation
terms (for instance, in classical electrodynamics radiation is modeled by
introducing a corresponding term proportional to the third derivation in
time of coordinates). Another argument for developing quantum field models
on the tangent bundle is the unclosed character of quantum electrodynamics
in which the renormalized amplitudes tend to $\infty $ with values of
momenta $p\rightarrow \infty $. This problem is avoided by introducing
additional suppositions, modification of fundamental principles and by
extending of the theory, which are less motivated from some physical points
of view. Similar problems, but more sophisticate, arises in modeling of
radiational dissipation in all variants of classical and quantum (super)
gravity and (supersymmetric) quantum field theories with higher derivations.
It is quite possible that to study physics of the Early Universe and propose
various scenarios of Kaluza-Klein compactification from higher dimensions to
the four dimensional one of the space-time it is more realistic to consider
theories with generic local anisotropy caused by fluctuations of quantum
high-dimensional space-time ''foam''. The above mentioned points to the
necessity to extend the geometric background of classical and quantum field
theories if a careful analysis of physical processes with non-negligible
beak reaction, quantum and statistical fluctuations, turbulence, random
dislocations and disclinations in continuous media.\

The bulk of this paper is devoted to a review of the necessary geometric
machinery for formulation of gravitational and matter field theories on
la-spaces, a presenting of the essential aspects (which has not so far been
published in the literature) of locally anisotropic strings and a discussion
of possible generation of la-gravity from string theory.

We use R.Miron and M.Anastasiei [10] conventions and basic results on the
geometry of la-spaces and la-gravity.

The plan of the paper is as follows. We begin with some models of la-spaces
in Sec. II\ , where we present the basic results on Finsler, Lagrange and
generalized Lagrange geometry, give an overview of the geometry of nonlinear
connections in vector bundles used as a general model of la-space and
propose a variant of locally anisotropic Einstein-Cartan theory. In Sec.
III\ we study the nonlinear $\sigma$-model and la-string propagation by
developing the d-covariant method of la-background field. Sec. IV\ is
devoted to problems of regularization and renormalization of the locally
anisotropic $\sigma$-model and a corresponding analysis of one- and two-loop
diagrams of this model. Scattering of la-gravitons and duality are
considered in Sec. V., and a summary and conclusions are drawn in Sec. VI.\

\section{Models of Locally Anisotropic Spaces}

We present a brief introduction into the geometry of vector and tangent
bundles (in brief, v-bundles and t-bundles) provided with nonlinear and
distinguished connections and metric structures [10].

\subsection{Finsler, Lagrange and Generalized Lagrange Spaces}

Let $M$ be a differentiable manifold of dimension $n,$ \ $\dim =n$
(differentiable means the class $C^\infty$ differentiability of functions)
and $TM$ is its tangent bundle.\ We denote by
$$
u=(x,y)=\{u^\alpha =(x^i,y^{(j)}=(x^i,y^j)\},%
$$
$(i,j,k,...=n-1),$ the local coordinates on $TM$ where $x=(x^i)$ are local
coordinates on $M$ and $y=(y^i)\;=(y^{(j)})$ are fiber coordinates.

A function $\Lambda :TM\rightarrow {\cal R}\ (\Lambda :(x,y)\rightarrow
\Lambda (x,y),{\cal R}\ $ is the real number field) is called a Finsler
metric on $M$ if it satisfies properties:

\begin{enumerate}
\item  $\Lambda (x,y\dot )>0,$ if $y\neq 0;$

\item  $\Lambda $ is a function of class $C^\infty $ for every $y\neq 0;$

\item  $\Lambda $ is of homogeneity $1$ on $y,$ i.e. $\Lambda (x,\zeta
y)=\zeta \Lambda (x,y),\zeta \in {\cal R\ ;\ }$

\item  the quadratic form
\end{enumerate}

\begin{equation}
g_{ij}(x,y)=\frac 12\frac{\partial ^2\Lambda ^2}{\partial y^i\partial y^j},%
\stackrel{}{\eqnum {2.1}}
\end{equation}
called the locally anisotropic metric tensor, la-metric, on ${\cal R}^n$ is
positive. Couple $F^n=(M,\Lambda (x,y))$ defines a Finsler space with the
fundamental function (metric) $\Lambda (x,y).$

The length of a curve $c:\gamma \in [0,1]\rightarrow x(\gamma )\in M$ on $%
F^n $ is given by the integral%
$$
s=\int_0^1\Lambda (x(\tau ),\frac{dx(\tau )}{d\tau })d\tau .
$$

We point out that Riemann spaces with metrics of type $g_{ij}(x)$ are a
particular class of Finsler spaces with the fundamental function of the form

$$
\Lambda (x,y)=\sqrt{g_{ij}(x)y^iy^j}
$$

La-metrics of type (2.1) where introduced by P.\ Finsler as a proposal of
generalization of Riemann geometry with the aim to some applications in
thermodynamics and anisotropic media physics. Variables $y=(y^j)$ can be
interpreted as parameters of local anisotropy of space-time or as
fluctuation parameters induced by classical and quantum field interactions
in locally anisotropic media with beak radiational reaction, turbulent and
stochastic processes.

The next step in extending the concept of la-space was the definition of
Lagrange spaces [15] used as a geometric background for non homogeneous
locally anisotropic processes [10]:

A Lagrangian on $M$ it is a differentiable function ${\cal L}:TM\rightarrow
{\cal R}$ ( locally defined as ${\cal L}:(x,y)\rightarrow {\cal L}(x,y)$ and
called the fundamental function ) with the property that fundamental tensor,
la-metric,
\begin{equation}
g_{ij}(x,y)=\frac 12\frac{\partial ^2{\cal L}}{\partial y^i\partial y^j}%
\eqnum{2.2}
\end{equation}
is non degenerate.

Couple $L^n=(M,{\cal L)}$ defines a Lagrange space.

It was also possible [10] to construct a consistent geometric theory for
spaces provided with a general fundamental tensor, la-metric (not obligatory
represented as a second order partial derivation of some fundamental
functions as in (2.1) or (2.2)).

Couple $M^n=(M,g_{ij}(x,y))$ with a general nondegenerate tensor $g_{ij}$
defines a generalized Lagrange space, GL-space.

In a more general form the geometry of la-spaces can be formulated in the
frame of vector and tangent bundles provided with nonlinear and
distinguished linear connections and metric structures

\subsection{Nonlinear Connection in Vector Bundles}

Let ${\cal E}=(E,\pi ,F,Gr,M)\;$ be a locally trivial v-bundle, where $F=%
{\cal R}^m$ is the typical vector space, $\dim =m,$ the structural group is
taken as $Gr=GL(m,{\cal R}),$ where $GL(m,{\cal R)}$ is the group of linear
transforms of ${\cal R}^m.$ We locally parameterize ${\cal E}$ by
coordinates $u^\alpha =(x^i,y^a),$ where $i,j,k,l,m,...,=0,1,...,n-1$ and $%
a,b,c,d,...=1,2,...,m.$

Coordinate transforms $(x^k,y^a)\rightarrow (x^{k^{\prime }},y^{a^{\prime
}}) $ on ${\cal E},$ considered as a differentiable manifold, are given by
formulas
\begin{equation}
x^{k^{\prime }}=x^{k^{\prime }}(x^k),y^{a^{\prime }}=M_a^{a^{\prime }}(x)y^a,%
\stackrel{}{\eqnum{2.3}}
\end{equation}
where $rank(\frac{\partial x^{k^{\prime }}}{\partial x^k})=n$ and $%
M_a^{a^{\prime }}(x)\in Gr.$

One of the fundamental objects in the geometry of la-spaces is the nonlinear
connection, in brief N-connection. The first global definition of
N-connection was given by W. Barthel [16] (a detailed study of N-connection
structures in v-bundles and basic references are contained in [10]). Here we
introduce the N-connection as a global decomposition of v-bundle ${\cal E}$
into horizontal, ${\cal HE},$ and vertical, ${\cal VE},$ subbundles of the
tangent bundle ${\cal TE}:$
\begin{equation}
{\cal TE}={\cal HE}\oplus {\cal VE}.{\eqnum{2.4}}
\end{equation}
With respect to a N-connection in ${\cal E}$ one defines a covariant
derivation operator
\begin{equation}
\nabla _YA=Y^i\left\{ \frac{\partial A^a}{\partial x^i}+N_i^a(x,A)\right\}
s_a,\stackrel{}{\eqnum{2.5}}
\end{equation}
where $s_a$ are local linear independent sections of ${\cal E},\Lambda
=\Lambda ^as_a$ and $Y=Y^is_i$ is the decomposition of vector field $Y$ on
local basis $s_i$ on $M.$ Differentiable functions $N_i^a(x,y)$ from (2.5)
(written as functions on $x^i$ and $y^a$) are called as coefficients
N-connection. We have these transformation laws for components $N_i^a$ under
coordinate transforms (2.3) :%
$$
N_{i^{\prime }}^{a^{\prime }}\frac{\partial x^{i^{\prime }}}{\partial x^i}%
=M_a^{a^{\prime }}N_i^a+\frac{\partial M_a^{a^{\prime }}}{\partial x^i}y^a .
$$
N-connection is also characterized by its curvature%
$$
\Omega =\frac 12{\Omega }_{ij}^adx^i\bigwedge dx^j\bigotimes \frac \partial
{\partial y^a},
$$
where $\bigwedge $ is the antisymmetric tensor product, with coefficients%
$$
{\Omega }_{ij}^a=\frac{\partial N_j^a}{\partial x^i}-\frac{\partial N_i^a}{%
\partial x^j}+N_j^b\frac{\partial N_i^a}{\partial y^b}-N_i^b\frac{\partial
N_j^a}{\partial y^b},
$$
and by its linearization defined as%
$$
\Gamma _{.bi\;}^a(x,y)=\frac{\partial N_i^a(x,y)}{\partial y^b}.
$$
The usual linear connections%
$$
\omega _{.b}^a=K_{.bi}^a(x)dx^i
$$
in v-bundle ${\cal E}$ form a particular class of N-connections with
coefficients parameterized as
$$
N_i^a(x,y)=K_{.bi}^a(x)y^b.
$$

\subsection{Distinguished Tensor Fields and Connections}

If in v-bundle ${\cal E}$ a N-connection structure is fixed, we must modify
the operation of partial derivation and to introduce a locally adapted, to
the N-connection, basis (frame)
\begin{equation}
\frac \delta {\delta u^\alpha }=(\frac \delta {\delta x^i}=\partial
_i-N_i^a(x,y)\frac \partial {\partial y^a},\frac \delta {\delta y^a}=\frac
\partial {\partial y^a}),\ {\eqnum{2.6}}
\end{equation}
instead of local coordinate basis
$$
\frac \partial {\partial u^a}=(\frac \partial {\partial x^i},\frac \partial
{\partial y^a}).
$$
The basis dual to $\frac \delta {\delta u^\alpha }$ is written as
\begin{equation}
\delta u^{\alpha} =(\delta x^i = dx^i, \delta y^a =dy^a + N_i^a(x,y)dx^i). {%
\eqnum{2.7}}
\end{equation}

By using bases(2.6) and (2.7) we can introduce the algebra of tensor
distinguished fields (d-fields, d-tensors) on ${\cal E}, {\cal C} = {\cal C}%
_{qs}^{pr},$ which is equivalent to the tensor algebra of the v-bundle $%
{\cal E}_d$ defined as
$$
\pi _d:{\cal HE}\oplus {\cal VE}\rightarrow {\cal TE}.
$$
An element $t\in {\cal C}_{qs}^{pr},$ d-tensor of type $\left(
\begin{array}{cc}
p & r \\
q & s
\end{array}
\right) $ , can be written in local form as%
$$
t=t_{j_1...j_qb_{1...}b_s}^{i_1...i_pa_1...a_r}(x,y)\frac \delta {\delta
x^{i_1}}\otimes ...\otimes \frac \delta {\delta x^{i_r}}\otimes
$$
$$
dx^{j_1}\otimes ...\otimes dx^{j_p}\otimes \frac \partial {\partial
y^{a_1}}\otimes ...\otimes \frac \partial {\partial y^{a_r}}\otimes \delta
y^{b_1}\otimes ...\otimes \delta y^{b_s}.
$$

In addition to d-tensors we can consider d-objects with differential
properties under group and coordinate transforms adapted to a global
splitting (2.4).

A distinguished linear connection, in brief d-con\-nec\-ti\-on, is defined
as a linear connection $D$ in \ ${\cal E}$ conserving as a parallelism the
Whitney sum \ ${\cal HE}\ \oplus \ {\cal VE}$ \ associated to a fixed
N-connection structure in ${\cal E}.$

Components $\Gamma _{.\beta \gamma }^\alpha $ of a d-connection $D$ are
introduced as
\begin{equation}
D_\gamma (\frac \delta {\delta u^\beta })=D_{(\frac \delta {\delta u^\gamma
})}(\frac \delta {\delta u^\beta })=\Gamma _{.\beta \gamma }^\alpha (\frac
\delta {\delta u^\alpha })\stackrel{}{. }\eqnum{2.8}
\end{equation}

We can define in a standard manner, with respect to locally adapted frame
(2.6), the components of torsion\thinspace $T_{.\beta \gamma}^\alpha $ and
curvature $R_{\beta .\gamma \tau }^{.\alpha }\,$ of d-connection $D:$
$$
T(\frac \delta {\delta u^\gamma },\frac \delta {\delta u^\beta })=%
$$
$$
D_{(\frac \delta {\delta u^\gamma })}\frac \delta {\delta u^\beta
}-D_{(\frac \delta {\delta u^\beta })}\frac \delta {\delta u^\gamma }-[\frac
\delta {\delta u^\gamma },\frac \delta {\delta u^\beta }]= T_{.\beta \gamma
}^\alpha \frac \delta {\delta u^\alpha },
$$
where
\begin{equation}
T_{.\beta \gamma }^\alpha =\Gamma _{.\beta \gamma }^\alpha -\Gamma _{.\gamma
\beta }^\alpha +w_{.\beta \gamma}^\alpha {\eqnum{2.9}}
\end{equation}
and
$$
R(\frac \delta {\delta u^\delta },\frac \delta {\delta u^\gamma },\frac
\delta {\delta u^\beta })=%
$$
$$
(D_{(\frac \delta {\delta u^\delta })}D_{(\frac \delta {\delta u^\gamma
})}-D_{(\frac \delta {\delta u^\gamma })}D_{(\frac \delta {\delta u^\delta
})}-D_{([\frac \delta {\delta u^\delta },\frac \delta {\delta u^\gamma
}])})\frac \delta {\delta u^\beta }=%
$$
$$
R_{\beta .\gamma \delta}^{.\alpha }\frac \delta {\delta u^\alpha },
$$
where%
$$
R_{\beta .\gamma \delta }^{.\alpha }=%
$$
\begin{equation}
\frac{\delta \Gamma _{.\beta \gamma}^\alpha } {\delta u^\delta }-\frac{%
\delta \Gamma _{.\beta \delta }^\alpha } {\delta u^\gamma }+ \Gamma _{.\beta
\gamma }^\varphi \Gamma _{.\varphi \delta }^\alpha -\Gamma _{.\beta \delta
}^\varphi \Gamma _{.\varphi \gamma }^\alpha +\Gamma _{.\beta \varphi
}^\alpha w_{.\gamma \delta }^\varphi .\stackrel{}{\eqnum{2.10}}
\end{equation}
In formulas (2.9) and (2.10) we have introduced nonholonomy coefficients $%
w_{.\beta \gamma }^\alpha $ of locally adapted frames (2.6):
\begin{equation}
[\frac \delta {\delta u^\alpha },\frac \delta {\delta u^\beta }]=\frac
\delta {\delta u^\alpha }\frac \delta {\delta u^\beta }-\frac \delta {\delta
u^\beta }\frac \delta {\delta u^\alpha }= w_{.\alpha \beta }^\gamma \frac
\delta {\delta u^\gamma } .\stackrel{}{\eqnum{2.11}}
\end{equation}

By straightforward calculations we can verify that distinguished torsion
(2.9) and curvature (2.10) satisfy Bianchi identities:%
$$
\sum_{[\alpha \beta \gamma ]}(D_\alpha R_{\mid \delta \mid .\beta \gamma
}^{.\varphi }+R_{\delta .\alpha \mid \psi \mid }^{.\varphi }T_{.\beta \gamma
}^\psi )=0,
$$
$$
\sum_{\left[ \alpha \beta \gamma \right] }(D_\alpha T_{.\beta \gamma
}^\delta +T_{.\alpha \beta }^\varphi T_{.\gamma \varphi }^\delta -R_{\alpha
.\beta \gamma }^{.\delta })=0,
$$
where $\sum_{\left[ \alpha \beta \gamma \right] }$ denotes the antisymmetric
cyclic summation on indices $\alpha ,\beta $ and $\gamma .$

The global decomposition (2.4) induces a corresponding invariant splitting
into horizontal $D_X^h=D_{hX}$ (h-derivation ) and vertical $D_X^v=D_{vX}$
(v-derivation) parts of the operator of covariant derivation $%
D,D_X=D_X^h+D_X^v,$ where $hX=X^i\frac \delta {\delta u^i}$ and $vX=X^a\frac
\partial {\partial y^a}$ are, respectively, the horizontal and vertical
components of vector field $X=hX+vX\,$ on ${\cal E}.$

Local coefficients $\left( L_{.jk}^i(x,y),L_{.bk}^a(x,y)\right) $ of
covariant h-derivation $D^h$ are introduced as
$$
D_{\left( \frac \delta {\delta x^k}\right) }^h\left( \frac \delta {\delta
x^j}\right) =L_{.jk}^i\left( x,y\right) \frac \delta {\delta x^i},%
$$
$$
D_{\left( \frac \delta {\delta x^k}\right) }^h\left( \frac \partial
{\partial y^b}\right) =L_{.bk}^a(x,y)\frac \partial {\partial y^a}
$$
and
\begin{equation}
D_{\left( \frac \delta {\delta x^k}\right) }^hf=\frac{\delta f}{\delta x^k}%
=\frac {\partial f}{\partial x^k}-N_k^a\left( x,y\right) \frac {\partial f}
{\partial y^a},\stackrel{}{\eqnum{2.12}}
\end{equation}
where $f\left( x,y\right) $ is a scalar function on ${\cal E}.$

Local co\-ef\-fi\-ci\-ents $\left( C_{.jk}^i\left( x,y\right)
,C_{.bk}^a\left( x,y\right) \right) $ of v-de\-ri\-va\-ti\-on $D^v$ are
intro\-du\-ced as
$$
D_{\left( \frac \partial {\partial y^c}\right) }^v\left( \frac \delta
{\delta x^j}\right) =C_{.jk}^i\left( x,y\right) \frac \delta {\delta x^i},%
$$
$$
D_{\left( \frac \partial {\partial y^c}\right) }^v\left( \frac \partial
{\partial y^b}\right) =C_{.bc}^a\left( x,y\right)
$$
and
\begin{equation}
D_{\left( \frac \partial {\partial y^c}\right) }^vf=\frac{\partial f}{%
\partial y^c}.\stackrel{}{\eqnum{2.13}}
\end{equation}

Let us consider, for example, the action of a covariant d-derivation on a
d-tensor field of type $\left(
\begin{array}{cc}
1 & 1 \\
1 & 1
\end{array}
\right) $
$$
Q=Q_{jb}^{ia}\frac \delta {\delta x^i}\otimes \frac \partial {\partial
y^a}\otimes dx^i\otimes \delta y^b.
$$
We write%
$$
D_XQ=D_X^hQ+D_X^vQ,
$$
where%
$$
D_X^hQ=X^kQ_{jb\mid k}^{ia}\frac \delta {\delta x^i}\otimes \frac \partial
{\partial y^a}\otimes dx^i\otimes \delta y^b,
$$
$$
Q_{ib\mid k}^{ia}=\frac{\delta Q_{jb}^{ia}}{\delta x^k}%
+L_{.hk}^iQ_{jb}^{ha}-L_{.jk}^hQ_{kb}^{ia}+L_{.ck}^aQ_{jb}^{ic}-L_{.bk}^cQ_{jc}^{ia},
$$
and%
$$
D_X^vQ=X^cQ_{jb\perp c}^{ia}\frac \delta {\delta x^i}\otimes \frac \partial
{\partial y^a}\otimes dx^i\otimes \delta y^b,
$$
$$
Q_{jb\perp c}^{ia}=\frac{\partial Q_{jb}^{ia}}{\partial y^c}%
+C_{.kc}^iQ_{jb}^{ka}-C_{.jc}^kQ_{kb}^{ia}+C_{.db}^aQ_{jb}^{id}-C_{.bc}^dQ_{jd}^{ia}.
$$

In order to computer the h- and v-components of torsion $T_{.\beta \gamma
}^\alpha $ and curvature $R_{\beta .\gamma \delta }^{.\alpha }$ we denote :%
$$
hT\left( \frac \delta {\delta x^k},\frac \delta {\delta x^j}\right)
=T_{.jk}^i\frac \delta {\delta x^k},hT\left( \frac \partial {\partial
y^b},\frac \delta {\delta x^i}\right) =P_{.jb}^i\frac \delta {\delta x^i},
$$
$$
vT\left( \frac \delta {\delta x^k},\frac \delta {\delta x^j}\right)
=T_{.jk}^a\frac \partial {\partial y^a},vT\left( \frac \partial {\partial
y^b},\frac \delta {\delta x^i}\right) =P_{.jb}^a\frac \partial {\partial
y^a},
$$
$$
vT\left( \frac \partial {\partial y^a},\frac \partial {\partial y^b}\right)
=S_{.bc}^a\frac \partial {\partial y^a}
$$
and\widetext
$$
R\left( \frac \delta {\delta x^k},\frac \delta {\delta x^j},\frac \delta
{\delta x^i}\right) =R_{l.jk}^{.i}\frac \delta {\delta x^i},R\left( \frac
\delta {\delta x^i},\frac \delta {\delta x^j},\frac \partial {\partial
y^a}\right) =R_{b.jk}^{.a}\frac \partial {\partial y^a},
$$
$$
R\left( \frac \partial {\partial y^c},\frac \delta {\delta x^k},\frac \delta
{\delta x^j}\right) =P_{j.kc}^{.i}\frac \delta {\delta x^i},R\left( \frac
\partial {\partial y^c},\frac \delta {\delta x^k},\frac \partial {\partial
y^b}\right) =P_{b.kc}^{.a}\frac \partial {\partial y^a},
$$
$$
R\left( \frac \partial {\partial y^c},\frac \partial {\partial y^b},\frac
\delta {\delta x^j}\right) =S_{j.bc}^{.i}\frac \delta {\delta x^i},R\left(
\frac \partial {\partial y^d},\frac \partial {\partial y^c},\frac \partial
{\partial y^b}\right) =S_{b.cd}^{.a}\frac \partial {\partial y^a}.
$$
Putting expressions (2.12) and(2.13) into (2.9) and (2.10), after
straightforward calculations, we obtain the h- and v-components of torsion,%
$$
T_{.jk}^i=T_{jk}^i,T_{ja}^i=C_{.ja}^i,T_{aj}^i=-C_{ja}^i,T_{.ja}^i=0,T_{.bc}^a=S_{.bc}^a,
$$
\begin{equation}
T_{.ij}^a=\frac{\delta N_i^a}{\delta x^j}-\frac{\delta N_j^a}{\delta x^i}%
,T_{.bi}^a=P_{.bi}^a=\frac{\partial N_i^a}{\partial y^b}%
-L_{.bj}^a,T_{.ib}^a=-P_{.bi,}^a\stackrel{}{\eqnum{2.14}}
\end{equation}
and of curvature%
$$
R_{h.jk}^{.i}=\frac{\delta L_{.hj}^i}{\delta x^h}-\frac{\delta L_{.hk}^i}{%
\delta x^j}+L_{.hj}^mL_{mk}^i-L_{.hk}^mL_{mj}^i+C_{.ha}^iR_{.jk}^a,
$$
$$
R_{b.jk}^{.a}=\frac{\delta L_{.bj}^a}{\delta x^k}-\frac{\delta L_{.bk}^a}{%
\delta x^j}+L_{.bj}^cL_{.ck}^a-L_{.bk}^cL_{.cj}^a+C_{.bc}^aR_{.jk}^c,
$$
$$
P_{j.ka}^{.i}=\frac{\partial L_{.jk}^i}{\partial y^k}-\left( \frac{\partial
C_{.ja}^i}{\partial x^k}%
+L_{.lk}^iC_{.ja}^l-L_{.jk}^lC_{.la}^i-L_{.ak}^cC_{.jc}^i\right)
+C_{.jb}^iP_{.ka}^b,
$$
$$
P_{b.ka}^{.c}=\frac{\partial L_{.bk}^c}{\partial y^a}-\left( \frac{\partial
C_{.ba}^c}{\partial x^k}+L_{.dk}^{c%
\,}C_{.ba}^d-L_{.bk}^dC_{.da}^c-L_{.ak}^dC_{.bd}^c\right)
+C_{.bd}^cP_{.ka}^d,
$$
$$
S_{j.bc}^{.i}=\frac{\partial C_{.jb}^i}{\partial y^c}-\frac{\partial
C_{.jc}^i}{\partial y^b}+C_{.jb}^hC_{.hc}^i-C_{.jc}^hC_{hb}^i,
$$
\begin{equation}
S_{b.cd}^{.a}=\frac{\partial C_{.bc}^a}{\partial y^d}-\frac{\partial
C_{.bd}^a}{\partial y^c}+C_{.bc}^eC_{.ed}^a-C_{.bd}^eC_{.ec}^a.\stackrel{}{^{%
\eqnum{2.15}}}
\end{equation}
\narrowtext

The components of the Ricci d-tensor
$$
R_{\alpha \beta }=R_{\alpha .\beta \tau }^{.\tau }
$$
with respect to locally adapted frame (2.7) are as follows:%
$$
R_{ij}=R_{i.jk}^{.k},R_{ia}=-^2P_{ia}=-P_{i.ka}^{.k},
$$
\begin{equation}
R_{ai}=^1P_{ai}=P_{a.ib}^{.b},R_{ab}=S_{a.bc}^{.c}.\stackrel{}{\eqnum{2.16}}
\end{equation}
We point out that because, in general, $^1P_{ai}\neq ^2P_{ia}$ the Ricci
d-tensor is non symmetric.

We shall also use the auxiliary torsionless d-connection $\tau _{.\beta
\gamma \text{ }}^\alpha $(defining the corresponding auxiliary covariant
d-derivation denoted as $\nabla $) introduced as to satisfy relations
\begin{equation}
\Gamma _{.\beta \tau }^\alpha =\tau _{.\beta \tau }^\alpha +T_{.\beta \tau
}^\alpha .{\eqnum{2.17}}
\end{equation}
The Riemann and Ricci d-tensors of the d-connection $\tau _{.\beta \gamma
}^\alpha $ are denoted respectively as $r_{\beta .\gamma \delta }^{.\alpha }$
and $r_{\alpha \beta }.$

\subsection{Distinguished Metric Structures in Vector Bundles}

Now, we shall analyze the compatibility conditions of N- and d-connection
and metric structures on v-bundle ${\cal E}.$

A metric field on ${\cal E},\ G\left( u\right) =G_{\alpha \beta }\left(
u\right) du^\alpha du^\beta ,$ is associated to a map%
$$
G\left( X,Y\right) :{\cal T}_u{\cal E}\times {\cal T}_u{\cal E}\rightarrow
R,
$$
parameterized by a non degenerate symmetric matrix
\begin{equation}
\left(
\begin{array}{cc}
\widehat{G}_{ij} & \widehat{G}_{ia} \\ \widehat{G}_{aj} & \widehat{G}_{ab}
\end{array}
\right) \stackrel{}{,}{\eqnum{2.18}}
\end{equation}
where
$$
\widehat{G}_{ij}=G\left( \frac \partial {\partial x^i},\frac \partial
{\partial x^j}\right) ,\widehat{G}_{ia}=G\left( \frac \partial {\partial
x^i},\frac \partial {\partial y^a}\right)
$$
and
$$
\widehat{G}_{ab}=G\left( \frac \partial {\partial y^a},\frac \partial
{\partial y^b}\right) .
$$

We choose a concordance between N-connection and G-metric structures by
imposing conditions%
$$
{G}\left( \frac \delta {\delta x^i},\frac \partial {\partial y^a}\right) =0,
$$
or, equivalently,
\begin{equation}
N_i^a\left( x,u\right) =\widehat{G}_{ib}\left( x,y\right) \widehat{G}%
^{ba}\left( x,y\right) ,\stackrel{}{\eqnum{2.19}}
\end{equation}
where $\widehat{G}^{ba}\left( x,y\right) $ are found as components of the
matrix $\widehat{G}^{\alpha \beta }$ being the inverse to $\widehat{G}%
_{\alpha \beta }.$ In this case metric ${G}$ on ${\cal E }$ is defined by
two independent d-tensors, $g_{ij}\left( x,y\right) $ of type $\left(
\begin{array}{cc}
2 & 0 \\
0 & 0
\end{array}
\right) $ and $h_{ab}\left( x,y\right) $ of type $\left(
\begin{array}{cc}
0 & 0 \\
0 & 2
\end{array}
\right) ,\,\,$ and with respect to locally adapted basis (2.7) can be
written as
$$
G\left( u\right) =G_{\alpha \beta }\left( u\right) \delta u^\alpha \delta
u^\beta =
$$
\begin{equation}
g_{ij}\left( x,y\right) dx^i\otimes dx^j+h_{ab}\left( x,y\right) \delta
y^a\otimes \delta y^b.\stackrel{}{\eqnum{2.20}}
\end{equation}

D-connection $\Gamma _{.\beta \gamma }^\alpha $ is compatible with d-metric
structure $G(u)$ on ${\cal E }$ if one holds equalities
\begin{equation}
D_\alpha G_{\beta \gamma }=0.\stackrel{}{\eqnum{2.21}}
\end{equation}

Having defined the d-metric (2.20) in ${\cal E}$ we can introduce the scalar
curvature of d-connection
\begin{equation}
{\overleftarrow{R}}=G^{\alpha \beta }R_{\alpha \beta }=R+S,{\eqnum{2.22}}
\end{equation}
where $R=g^{ij}R_{ij}$ and $S=h^{ab}S_{ab}.$ The scalar curvature of the
auxiliary torsionless d-connection (2.17) can be written in a similar
manner, $\overleftarrow{r}=G^{\alpha \beta }r_{\alpha \beta .}$

Both d-metric (2.20) and N-connection on ${\cal E}$ define the so-called
canonic d-connection, $C\Gamma \left( N\right) -$connection [10], with
lo\-cally adapted com\-po\-nents
$$
^{\circ }\Gamma _{.jk}^i=\left( ^{\circ }L_{.jk}^i,^{\circ
}L_{.bi}^a,^{\circ }C_{.jk}^i,^{\circ }C_{.bc}^a\right)%
$$
expressed as
$$
^{\circ }L_{.jk}^i=\frac 12g^{ip}\left( \frac{\delta g_{pj}}{\delta x^k}+
\frac{\delta g_{pk}}{\delta x^j}-\frac{\delta g_{jk}}{\delta x^k}\right) ,
$$
$$
^{\circ }L_{.bi}^a=\frac{\partial N_i^a}{\partial y^b}+\frac 12h^{ac}\left(
\frac{\delta h_{bc}}{\delta x^i}-\frac{\partial N_i^d}{\partial y^b}h_{dc}-
\frac{\partial N_i^d}{\partial y^c}h_{db}\right) ,
$$
$$
^{\circ }C_{.jb}^i=\frac 12g^{ik}\frac{\partial g_{jk}}{\partial y^b},
$$
$$
^{\circ }C_{.bc}^a=\frac 12h^{ad}\left( \frac{\partial h_{dc}}{\partial y^b}%
+ \frac{\partial h_{db}}{\partial y^c}-\frac{\partial h_{bc}}{\partial y^d}%
\right) .
$$

Putting components of $C\Gamma \left( N\right) -$connection $^{\circ }\Gamma
_{.\beta \gamma }^\alpha $ instead of components of arbitrary d-connection $%
\Gamma _{.\beta \gamma }^\alpha $ into formulas (2.14),(2.15) and (2.16) we
can calculate respectively the adapted coefficients of the canonic torsion $%
^{\circ }T_{.\beta \gamma }^\alpha ,$ curvature $^{\circ }R_{\beta .\gamma
\delta }^{.\alpha },$ and Ricci d-tensor $^{\circ }R_{\alpha \beta }.$

\subsection{The Almost Hermitian Model of Generalized Lagrange Spaces}

Let us fix a N-connection in tangent bundle $\left( TM,\tau ,M\right) $ and
consider a GL-space $M^n=\left( M,g_{ij}\left( x,y\right) \right) .\,\,$ We
note that all formulas presented in the previous subsections hold good on $%
TM.$ In this case the dimension of the structural fiber is equal to the
dimension of base space. Both type of indices (on the base, i,j,k,..., and
on the typical fiber, (i)=i,(j)=j,(k)=k,...) run from 0 to n-1. Every
compatible with metric d-connection $L\Gamma \left( N\right) =\left(
\overline{L}_{.jk}^i,\overline{C}_{jk}^i\right) $ on $M^n\,$ can be written
with respect to a locally adapted basis as
$$
\overline{L}_{.jk}^i=L_{.jk}^i+\frac 12g^{ir}\left(
g_{rm}T_{.jk}^m-g_{jm}T_{.rk}^m+g_{km}T_{.jr}^m\right) ,
$$
\begin{equation}
\overline{C}_{.jk}^i=C_{.jk}^i+\frac 12g^{ir}\left(
g_{im}S_{.jk}^m-g_{jm}S_{.rk}^m+g_{km\,}S_{.jr}^m\right) \stackrel{}{%
\eqnum{2.23}}
\end{equation}
with ar\-bi\-trary pre\-scribed tor\-sion ten\-sors
$$
T_{.jk}^i=-T_{.kj}^i\ , \mbox{and} \ S_{.ij}^i=-S_{.kj}^i.\,%
$$
Co\-ef\-fi\-ci\-ents%
$$
L_{.jk}^i\left( x,y\right) =\frac 12g^{ir}\left( \frac{\delta g_{jk}}{\delta
x^k}+\frac{\delta g_{kr}}{\delta x^j}-\frac{\delta g_{jk}}{\delta x^r}%
\right) ,
$$
\begin{equation}
C_{.jk}^i\left( x,y\right) =\frac 12g^{ir}\left( \frac{\partial g_{jr}}{%
\partial y^k}+\frac{\partial g_{kr}}{\partial y^j}-\frac{\partial g_{jk}}{%
\partial y^r}\right) \stackrel{}{\eqnum{2.24}}
\end{equation}
form the unique torsionless canonic connection $C\Gamma \left( N\right)
=\left( L_{.jk}^{i\,},C_{.jk}^i\right) $ on $M^n.$

N-connection defines naturally an almost complex structure [10] $F\,$ on $%
TM\,$ acting as%
$$
F\left( \frac \delta {\delta x^i}\right) =-\frac \partial {\partial y^i},
$$
and%
$$
F\left( \frac \partial {\partial y^i}\right) =\frac \delta {\delta x^i}.
$$
If an almost tangent structure $J\,\,$ , with the property that%
$$
J\left( \frac \delta {\delta x^i}\right) =\frac \delta {\delta x^i}
$$
and%
$$
J\left( \frac \partial {\partial y^i}\right) =-\frac \partial {\partial
y^i},
$$
and a d-connection $D$ satisfying condition $DJ=0\,$ are defined on $TM,$
one calls $D\,$ the normal connection on \thinspace $M^n.$

One says that a linear connection $D\,$ on $TM$ is the N- lift of a normal
d-connection on $M$ if the compatibility of the almost complex structure $F$
and of d-connection structure holds, i.e. if $DF=0.$

Condition $g\left( FX,FY\right) =g\left( X,Y\right) ,$ where $X\,$ and $Y$
are arbitrary d-vector fields, for d-metric (2.20) on $TM$ implies that $%
g_{ij}\left( x,y\right) =g_{\left( i\right) \left( j\right) }\left(
x,y\right) ,$ i.e. the N-lift on $TM$ of the fundamental tensor $%
g_{ij}\left( x,y\right) $ on $M$ defines a covariant second order symmetric
non degenerate d-field
\begin{equation}
\tensor{G}\left( x,y\right) =g_{ij}\left( x,y\right) dx^idx^j+g_{ij}\left(
x,y\right) \delta y^i\delta y^j.\stackrel{}{\eqnum{2.25}}
\end{equation}
Let call this d-metric as the H$^{2n}$-metric.\ By straightforward
calculations we can verify that H$^{2n}$-metric is compatible with the
N-lift of metric d-connection $L\Gamma \left( N\right) $ (2.23), i.e.
\begin{equation}
\overline{D}_\gamma \tensor{G}_{\alpha \beta }=0.\stackrel{}{\eqnum{2.26}}
\end{equation}

Components of connection $\overline{D}$ on $TM$ with respect to a locally
adapted basis are computed [10] as%
$$
\overline{\Gamma ^i}_{jk}=\overline{L}_{.jk}^i,\overline{\Gamma ^i}_{.j(k)}=
\overline{C}_{.jk}^i,
$$
\begin{equation}
\overline{\Gamma ^{(i)}}_{(j)k}=\overline{L}_{.jk}^i,\overline{\Gamma ^{(i)}}%
_{(j)(k)}=\overline{C}_{.jk}^i {\eqnum{2.27}}
\end{equation}

Space $H^{2n}=\left( TM,G,F\right) ,$ in brief $H^{2n}$-space, defines the
almost Hermitian model of GL-space $M^n=(M,g_{ij}(x,y)).$

Components of torsion $\overline{T^\alpha }_{.\beta \gamma },$ curvature and
Ricci tensor $\overline{R}_{\beta .\gamma \delta }^{.\alpha }$ and Ricci
tensor $\overline{R}_{\alpha \beta }$ can be computed respectively by
putting components (2.27) into formulas (2.14),(2.15) and (2.16) considered
on tangent bundle $TM.$

Finally, in this subsection, we remark that we can consider N-lifts on $TM$
of a Finsler (or Lagrange) la-metric, if instead of coefficients of
GL-metric $g_{ij}$ the coefficients (2.1) (on (2.2)) are used.

\subsection{Einstein-Cartan Equations on Locally Anisotropic Spaces}

Now we shall analyze the problem of formulation of models of gravitational
interactions on la-spaces and propose a version of la-gravity theory with
torsion on such spaces.

There are developed some approaches to la-gravity (see details and
discussions of physical aspects in Refs [11]). Here we emphasize that the
Einstein like theory on $H^{2n}$ - spaces [10] extends the general
relativity theory on la-spaces in a manner as to have compatible nonlinear
and distinguished linear connections and metric structures (see (2.26)).
This property allows a straightforward generalization of classical and
quantum field theories on (pseudo)- Riemannian spaces to similar ones on H$%
^{2n}$-spaces by changing respectively partial derivations $\frac \partial
{\partial x^i},$ metric $g_{ij}\left( x\right) $ and metric connection $%
\Gamma _{.jk}^i\left( x\right) $ into locally adapted operators $\frac
\delta {\delta u^\alpha }=\left( \frac \delta {\delta x^i},\frac \partial
{\partial y^j}\right) $ (see (2.6)), H$^{2n}$-metric $\tensor{G}_{\mu \nu
}\left( u\right) $ (see (2.25)) and H$^{2n}$-connection $\overline{\Gamma
^\alpha }_{.\beta \gamma }$ (see (2.27)). We also note that on generic H$%
^{2n}$-spaces the locally adapted to N-connection bases are nonholonomic,
torsion is not vanishing and the Ricci d-tensor is not symmetric.

As a general model of la-space we shall use a v-bundle space ${\cal E}$
provided with N-connection, d-connection and d-metric satisfying compatibility
conditions (2.21).

Einstein-Cartan equations for d-connection $\Gamma _{.\beta \gamma }^\alpha $
are written in a standard manner by using the torsion d-tensor (2.6), the
Ricci d-tensor (2.16), d-metric (2.20) and the scalar curvature (2.22):
\begin{equation}
R_{\alpha \beta }-\frac 12G_{\alpha \beta }{\overleftarrow R}+ \widetilde{%
\lambda }G_{\alpha \beta }=\kappa _{\left( 1\right) }T_{\alpha \beta },%
\stackrel{}{\eqnum{2.28}}
\end{equation}
\begin{equation}
T_{.\beta \gamma }^\alpha +\delta _\beta ^\alpha T_{.\gamma \tau }^\tau
-\delta _\gamma ^\alpha T_{.\beta \tau }^\tau =\kappa _{\left( 2\right)
}Q_{.\beta \gamma }^\alpha ,\stackrel{}{\eqnum{2.29}}
\end{equation}
where $\kappa _{\left( 1\right) }$and $\kappa _{\left( 2\right) }$ are
interaction constants. In (2.28) we have introduced the locally anisotropic
cosmological term $\widetilde{\lambda }G_{\alpha \beta }$ and
energy-momentum d-tensor $T_{\alpha \beta }.$ D-tensor $Q_{.\beta \gamma
}^\alpha $ from (2.29) is interpreted as the locally anisotropic spin-matter
source of torsion.

R. Miron and M. Anastasiei [17,10] considered on v-bundle $\xi $ the
Einstein equations with prescribed torsion and N-connection and without
cosmological term (we present here the h- and v-projections of theirs
equations, see, correspondingly, the necessary h- and v-components of
d-metric, Ricci d-tensor and scalar curvature in (2.20),(2.16) and (2.22)):%
$$
R_{ij}-\frac 12\left( R+S\right) g_{ij}=\kappa _{\left( 1\right) }T_{ij},
$$
$$
S_{ab}-\frac 12\left( R+S\right) h_{ab}=\kappa _{\left( 2\right) }T_{ab,}
$$
$$
^1P_{ai}=\kappa _{\left( 1\right) }T_{ai,}
$$
\begin{equation}
^2P_{ia}=-\kappa _{\left( 2\right) }T_{ia.}\stackrel{}{\eqnum{2.30}}
\end{equation}

La-gravitational field equa\-tions (2.30) con\-tain $(n+m)^2$ equa\-ti\-ons
for
$$
nm+\frac 12[n(n+1)+m(m+1)+n^2(n-1)+m^2(m-1)]
$$
unknown variables $N_i^a,g_{ij},h_{ab},T_{.jk}^i$ and $S_{.bc}^a.$ The last
number is greater than the first one for $n\geq 3$ and $m\geq 1.$ We have
add equations (2.29) in order to obtain a closed system of equations with
metric-N-connection constraints of type (2.18). The introduction of the
cosmological term presents interest for a further study of possible
cosmological consequences of la-gravity. In [13] we formulated a more
general variant of gauge la-gravity with dynamical equations for torsion
instead of algebraic equations. In this work we shall consider both the
Einstein like field equations and the equations for the curvature of
N-connection and torsion of d-connection arising as constraints for a
consistent propagation of strings on la-backgrounds.

Finally, in this section, we note that la-gravitational field equations
(2.30) will be transformed into Einstein ones for multidimensional gravity
if the nonlinear connection and torsion of linear conection vanish. Usualy
one considers different scenarious of compatification from higher dimensions
to four dimensional one by intoducing by hand or postulating a kind of
spontaneous symmetry broking  of local multidimensional gravitational symmetry
By using the N-connection we can modelate dynamical variants of "splitting"
of space-time dimensions and consider physical consequences for possible
relict effects of locally anisotropic fluctuations.

\section{LOCALLY ANISOTROPIC STRINGS AND SIGMA-MODELS}

In this section we present a generalization of some necessary results on
nonlinear $\sigma $-model and string propagation to the case of
la-backgrounds. Calculations on both type of locally anisotropic and
isotropic spaces are rather similar if we accept the Miron and Anastasiei
[10] geometric formalism. We emphasize that on la-backgrounds we have to
take into account the distinguished character, by N-connection, of geometric
objects.

\subsection{Action of Nonlinear $\sigma $-model and Torsion of La-space}

Let a map of a two-dimensional (2d), for simplicity, flat space $M^2$ into
la-space $\xi $ defines a $\sigma $ -model field $u^\mu \left( z\right)
=\left( x^i\left( z\right) ,y^a\left( z\right) \right) ,$ where $z=\left\{
z^A,A=0,1\right\} $ are two-dimensional complex coordinates on $M^2.$ The
moving of a bosonic string in la-space is governed by the nonlinear $\sigma $%
-model action (see, for instance, [1,13,4] for details on locally isotropic
spaces):
$$
I=\frac 1{\lambda ^2}\int d^2z[\frac 12\sqrt{\gamma }\gamma ^{AB}\partial
_Au^\mu \left( z\right) \partial _Bu^\nu \left( z\right) G_{\mu \nu }\left(
u\right) +
$$
\begin{equation}
\frac{\widetilde{n}}3\epsilon ^{AB}\partial _Au^\mu \partial _B u^\nu b_{\mu
\nu }\left( u\right) +\frac{\lambda ^2}{4\pi }\sqrt{\gamma }R^{(2)}\Phi
\left( u\right) ],\stackrel{}{\eqnum{3.1}}
\end{equation}
where $\lambda ^2$ and $\widetilde{n}\,$ are interaction constants, $\Phi
\left( u\right) $ is the dilaton field, $R^{(2)}$ is the curvature of the 2d
world sheet provided with metric $\gamma _{AB},\gamma =\det \left( \gamma
_{AB}\right) $ and $\partial _A=\frac \partial {\partial z^A},$ tensor $%
\epsilon ^{AB}$ and d-tensor $b_{\mu \nu }$ are antisymmetric.

From the viewpoint of string theory we can interpret $b_{\alpha \beta }$ as
the vacuum expectation of the antisymmetric, in our case locally
anisotropic, d-tensor gauge field $B_{\alpha \beta \gamma }$ (see
considerations for locally isotropic models in [18,19] and the
Wess-Zumino-Witten model [20,21], which lead to the conclusion [22] that $
\widetilde{n}$ takes only integer values and that in the perturbative
quantum field theory the effective quantum action depends only on B$_{...}$
and does not depend on b$_{...}$ ).

In order to obtain compatible with N-connection motions of la-strings we
consider these relations between d-tensor $b_{\alpha \beta },$ strength $%
B_{\alpha \beta \gamma }=\delta _{[\alpha }b_{\beta \gamma ]}$ and torsion $%
T_{.\beta \gamma }^\alpha :$%
\begin{equation}
\delta _\alpha b_{\beta \gamma }=T_{\alpha \beta \gamma },\stackrel{}{%
\eqnum{3.2}}
\end{equation}
with the integrability conditions
\begin{equation}
\Omega _{ij}^a\partial _a b_{\beta \gamma} = \delta _{[i}T_{j]\beta \gamma },%
\stackrel{}{\eqnum{3.3}}
\end{equation}
where $\Omega _{ij}^a$ are the coefficients of the N-connection curvature.
In this case we can express $B_{\alpha \beta \gamma }=T_{[\alpha \beta
\gamma ]}.$ Conditions (3.2) and (3.3) define a simplified model of
la-strings when the $\sigma $-model antisymmetric strength is induced from
the la-background torsion. More general constructions are possible by using
normal coordinates adapted to both N-connection and torsion structures on
la-background space. For simplicity, we omit such considerations in this
work.

Choosing the complex (conformal) coordinates $z=\iota ^0+i\iota ^1,\overline{%
z}=\iota ^0-i\iota ^1,$ where $\iota ^A,A=0,1$ are real coordinates, on the
world sheet we can represent the two-dimensional metric in the conformally
flat form:
\begin{equation}
ds^2=e^{2\varphi }dzd\overline{z},\stackrel{}{\eqnum{3.4}}
\end{equation}
where $\gamma _{z\overline{z}}=\frac 12e^{2\varphi }$ and $\gamma
_{zz}=\gamma _{\overline{z}\overline{z}}=0.$

Let us consider an la-field $U\left( u\right) ,u\in \xi $ taking values in $%
{\cal G}$ being the Lie algebra of a compact and semi simple Lie group,%
$$
U\left( u\right) =\exp [i\varphi \left( u\right) ],\varphi \left( u\right)
=\varphi ^{\underline{\alpha }}\left( u\right) q^{\underline{\alpha }},
$$
where $q^{\underline{\alpha }}$ are generators of the Lie algebra with
antisymmetric structural constants $f^{\underline{\alpha }\underline{\beta }
\underline{\gamma }}$ satisfying conditions%
$$
[q^{\underline{\alpha }},q^{\underline{\beta }}]=2if^{\underline{\alpha }
\underline{\beta }\underline{\gamma }}q^{\underline{\gamma }},\quad tr(q^{
\underline{\alpha }}q^{\underline{\beta }})=2\delta ^{\underline{\alpha }
\underline{\beta }}.
$$

The action of the Wess-Zumino-Witten type la-model should be written as
\begin{equation}
I\left( U\right) =\frac 1{4\lambda ^2}\int d^2z\ tr\left( \partial
_AU\partial ^AU^{-1}\right) +\widetilde{n}\Gamma \left[ U\right] ,\stackrel{%
}{\eqnum{3.5}}
\end{equation}
where $\Gamma \left[ U\right] $ is the standard topologically invariant
functional [22]. For perturbative calculations in the framework of the model
(3.1) it is enough to know that as a matter of principle we can represent
the action of our theory as (3.5) and to use d-curvature $r_{\beta .\gamma
\delta }^{.\alpha }$ for the torsionless d-connection $\tau _{.\beta \gamma
,}^\alpha $ see (2.17), and strength $B_{\alpha \beta \gamma }$ respectively
expressed as
$$
R_{\alpha \beta \gamma \delta }=f_{\underline{\alpha }\underline{\beta }
\underline{\tau }}f_{\underline{\gamma }\underline{\delta }\underline{\tau }%
}V_\alpha ^{\underline{\alpha }}V_\beta ^{\underline{\beta }}V_\gamma ^{
\underline{\gamma }}V_\delta ^{\underline{\delta }}
$$
and%
$$
B_{\alpha \beta \gamma }=\eta f_{\underline{\alpha }\underline{\beta }
\underline{\tau }}V_\alpha ^{\underline{\alpha }}V_\beta ^{\underline{\beta }%
}V_\tau ^{\underline{\tau }},
$$
where a new interaction constant $\eta \equiv \frac{\widetilde{n}\lambda ^2}{%
2\pi }$ is used and $V_\alpha ^{\underline{\alpha }}$ is a locally adapted
vielbein associated to the metric (2.20):%
$$
G_{\alpha \beta }=V_\alpha ^{\underline{\alpha }}V_\beta ^{\underline{\beta }%
}\delta ^{\underline{\alpha }\underline{\beta }}
$$
and
\begin{equation}
G^{\alpha \beta }V_\alpha ^{\underline{\alpha }}V_\beta ^{\underline{\beta }%
}=\delta ^{\underline{\alpha }\underline{\beta }}.\stackrel{}{\eqnum{3.6}}
\end{equation}
For simplicity, we shall omit underlining of indices if this will not give
rise to ambiguities.

Finally, in this subsection, we remark that for $\eta =1$ we obtain a
conformally invariant two-dimensional quantum field theory (being similar to
those developed in [23,20]).

\subsection{The D-covariant Method of La-background Field and $\sigma $%
-models}

Suggesting the compensation of all anomalies we can fix the gauge for the
two-dimensional metric when action (3.1) is written as
\begin{equation}
I\left[ u\right] =\frac 1{2\lambda ^2}\int d^2z\{G_{\alpha \beta }\eta
^{AB}+\frac 23b_{\alpha \beta }\epsilon ^{AB}\}\partial _Au^\alpha \partial
_Bu^\beta ,\stackrel{}{\eqnum{3.1a}}
\end{equation}
where $\eta ^{AB}$ and $\epsilon ^{AB\text{ }}$ are, respectively, constant
two-dimensional metric and antisymmetric tensor. The covariant method of
background field, as general references see [24-27], can be extended for
la-spaces. Let consider a curve in $\xi $ parameterized as $\rho ^\alpha
\left( z,s\right) ,s\in [0,1],$ satisfying autoparallel equations%
$$
\frac{d^2\rho ^\alpha \left( z,s\right) }{ds^2}+\Gamma _{.\beta \gamma
}^\alpha \left[ \rho \right] \frac{d\rho ^\beta }{ds}\frac{d\rho ^\gamma }{ds%
}=
$$
$$
\frac{d^2\rho ^\alpha \left( z,s\right) }{ds^2}+\tau _{.\beta \gamma
}^\alpha \left[ \rho \right] \frac{d\rho ^\beta }{ds}\frac{d\rho ^\gamma }{ds%
}=0,
$$
with boundary con\-di\-tions
$$
\rho \left( z,s=0\right) =u\left( z\right) \ \mbox{and} \ \rho \left(
z,s=1\right) u\left( z\right) +v\left( z\right) .%
$$
For simplicity, hereafter we shall consider that d-con\-nec\-ti\-on ${{\Gamma%
}^{\alpha}}_{\beta\gamma}$ is defined by d-metric (2.20) and
N-con\-nec\-ti\-on structures, i.e.
$$
{{\Gamma}^{\alpha}}_{\beta \gamma}  ={}^{\circ}{{\Gamma}^{\alpha}}_{\beta
\gamma} .%
$$
The tangent d-vector $\zeta ^\alpha =\frac d{ds}\rho ^\alpha ,$ where $\zeta
^\alpha \mid _{s=0}=\zeta _{\left( 0\right) }^\alpha $ is chosen as the
quantum d-field. Then the expansion of action $I[u+v\left( \zeta \right) ],$
see (3.1a), as power series on $\zeta ,$
$$
I[u+v\left( \zeta \right) ]=\sum_{k=0}^\infty I_k,
$$
where
$$
I_k=\frac 1{k!}\frac{d^k}{ds^k}I\left[ \rho \left( s\right) \right] \mid
_{s=0},
$$
defines d-covariant, depending on the background d-field, interaction
vortexes of locally anisotropic $\sigma $-model.

In order to compute $I_k\,$ it is useful to consider relations%
$$
\frac d{ds}\partial _A\rho ^\alpha =\partial _A\zeta ^\alpha ,\frac
d{ds}G_{\alpha \beta }=\zeta ^\tau \delta _\tau G_{\alpha \beta },
$$
$$
\partial _AG_{\alpha \beta }=\partial _A\rho ^\tau \delta _\tau G_{\alpha
\beta },
$$
to introduce auxiliary operators%
$$
\left( \widehat{\nabla }_A\zeta \right) ^\alpha =\left( \nabla _A\zeta
\right) ^\alpha -G^{\alpha \tau }T_{\left[ \alpha \beta \gamma \right]
}\left[ \rho \right] \epsilon _{AB}\partial ^B\rho ^\gamma \zeta ^\beta ,
$$
$$
\left( \nabla _A\zeta \right) ^\alpha =\left[ \delta _\beta ^\alpha \partial
_A+\tau _{.\beta \gamma }^\alpha \partial _A\rho ^\gamma \right] \zeta
^\beta ,
$$
\begin{equation}
\nabla \left( s\right) \xi ^\lambda =\zeta ^\alpha \nabla _\alpha \xi
^\lambda =\frac d{ds}\xi ^\lambda +\tau _{.\beta \gamma }^\lambda \left[
\rho \left( s\right) \right] \zeta ^\beta \xi ^\gamma ,\stackrel{}{%
\eqnum{3.7}}
\end{equation}
having properties%
$$
\nabla \left( s\right) \zeta ^\alpha =0,\nabla \left( s\right) \partial
_A\rho ^\alpha =\left( \nabla _A\zeta \right) ^\alpha ,
$$
$$
\nabla ^2\left( s\right) \partial _A\rho ^\alpha =r_{\beta .\gamma \delta
}^{.\alpha }\zeta ^\beta \zeta ^\gamma \partial _A\rho ^\delta ,
$$
and to use the curvature d-tensor of d-connection (3.7), \widetext
$$
\widehat{r}_{\beta \alpha \gamma \delta }=r_{\beta \alpha \gamma \delta
}-\nabla _\gamma T_{\left[ \alpha \beta \delta \right] }+\nabla _\delta
T_{\left[ \alpha \beta \gamma \right] }-T_{\left[ \tau \alpha \gamma \right]
}G^{\tau \lambda }T_{\left[ \lambda \delta \beta \right] }+T_{\left[ \tau
\alpha \delta \right] }G^{\tau \lambda }T_{\left[ \lambda \gamma \beta
\right] }.
$$

Values $I_k$ can be computed in a similar manner as in [19,22,27], but in
our case by using corresponding d-connections and d-objects. Here we present
the first four terms in explicit form:%
$$
I_1=\frac 1{2\lambda ^2}\int d^2z2G_{\alpha \beta }\left( \widehat{\nabla }%
_A\zeta \right) ^\alpha \partial ^Au^\beta ,
$$
$$
I_2=\frac 1{2\lambda ^2}\int d^2z\{\left( \widehat{\nabla }_A\zeta \right)
^2+\widehat{r}_{\beta \alpha \gamma \delta }\zeta ^\beta \zeta ^\gamma
\left( \eta ^{AB}-\epsilon ^{AB}\right) \partial _Au^\alpha \partial
_Bu^\beta \},
$$
$$
I_3=\frac 1{2\lambda ^2}\int d^2z\{\frac 43(r_{\beta \alpha \gamma \delta
}-G^{\tau \epsilon }T_{\left[ \epsilon \alpha \beta \right] }T_{\left[ \tau
\gamma \delta \right] })\partial _Au^\alpha (\widehat{\nabla }^A\zeta
^\delta )\zeta ^\beta \zeta ^\gamma +\frac 43\nabla _\alpha T_{\left[ \delta
\beta \gamma \right] }\partial _Au^\beta \epsilon ^{AB}(\widehat{\nabla }%
_B\zeta ^\gamma )\zeta ^\alpha \zeta ^\delta +
$$
$$
\frac 23T_{\left[ \alpha \beta \gamma \right] }(\widehat{\nabla }_A\zeta
^\alpha )\epsilon ^{AB}(\widehat{\nabla }\zeta ^\beta )\zeta ^\gamma +\frac
13\left( \nabla _\lambda r_{\beta \alpha \gamma \delta }+4G^{\tau \epsilon
}T_{\left[ \epsilon \lambda \alpha \right] }\nabla _\beta T_{\left[ \gamma
\delta \tau \right] }\right) \partial _Au^\alpha \partial ^Au^\delta \zeta
^\gamma \zeta ^\lambda +
$$
$$
\frac 13\left( \nabla _\alpha \nabla _\beta T_{\left[ \tau \gamma \delta
\right] }+2G^{\lambda \epsilon }G^{\varphi \phi }T_{\left[ \alpha \lambda
\varphi \right] }T_{[\epsilon \beta \delta ]}T_{\left[ \phi \tau \gamma
\right] }+2r_{\alpha .\beta \gamma }^{.\lambda }T_{\left[ \alpha \tau \delta
\right] }\right) \partial _Au^\gamma \epsilon ^{AB}\left( \partial _B\zeta
^\tau \right) \zeta ^\alpha \zeta ^\beta \},
$$
$$
I_4=\frac 1{4\lambda ^2}\int d^2z\{\left( \frac 12\nabla _\alpha r_{\gamma
\beta \delta \tau }-G^{\lambda \epsilon }T_{\left[ \epsilon \beta \gamma
\right] }\nabla _\alpha T_{\left[ \lambda \delta \tau \right] }\right)
\partial _Au^\beta (\widehat{\nabla }^A\zeta ^\tau )\zeta ^\alpha \zeta
^\gamma \zeta ^\delta +
$$
$$
\frac 13r_{\beta \alpha \gamma \delta }(\widehat{\nabla }_A\zeta ^\alpha )(
\widehat{\nabla }^A\zeta ^\delta )\zeta ^\beta \zeta ^\gamma +(\frac
1{12}\nabla _\alpha \nabla _\beta r_{\delta \gamma \tau \lambda }+\frac
13r_{\delta .\tau \gamma }^{.\kappa }r_{\beta \kappa \alpha \lambda }-\frac
12(\nabla _\alpha \nabla _\beta T_{\left[ \gamma \tau \epsilon \right]
})G^{\epsilon \pi }T_{\left[ \pi \delta \lambda \right] }-
$$
$$
\frac 12r_{\alpha .\beta \gamma }^{.\kappa }T_{\left[ \kappa \tau \epsilon
\right] }G^{\epsilon \pi }T_{\left[ \pi \delta \lambda \right] }+\frac
16r_{\alpha .\beta \epsilon }^{.\kappa }T_{\left[ \kappa \gamma \tau \right]
}G^{\epsilon \pi }T_{\left[ \epsilon \delta \lambda \right] })\partial
_Au^\gamma \partial ^Au^\lambda \zeta ^\alpha \zeta ^\beta \zeta ^\delta
\zeta ^\tau +[\frac 1{12}\nabla _\alpha \nabla _\beta \nabla _\gamma
T_{\left[ \lambda \delta \tau \right] }+
$$
$$
\frac 12\nabla _\alpha \left( G^{\kappa \epsilon }T_{\left[ \epsilon \lambda
\delta \right] }\right) r_{\beta \kappa \gamma \tau }+\frac 12\left( \nabla
_\alpha T_{\left[ \pi \beta \kappa \right] }\right) G^{\pi \epsilon
}G^{\kappa \nu }T_{\left[ \epsilon \gamma \delta \right] }T_{\left[ \nu
\lambda \tau \right] }-\frac 13G^{\kappa \epsilon }T_{\left[ \epsilon
\lambda \delta \right] }\nabla _\alpha r_{\beta \kappa \delta \tau }]\times
$$
$$
\partial _Au^\delta \epsilon ^{AB}\partial _Bu^\tau \zeta ^\lambda \zeta
^\alpha \zeta ^\beta \zeta ^\gamma +\frac 12\left[ \nabla _\alpha \nabla
_\beta T_{\left[ \tau \gamma \delta \right] }+T_{\left[ \kappa \delta \tau
\right] }r_{\alpha .\beta \gamma }^{.\kappa }+r_{\alpha .\beta \delta
}^{.\kappa }T_{\left[ \kappa \gamma \tau \right] }\right] \times
$$
\begin{equation}
\partial _Au^\gamma \epsilon ^{AB}(\widehat{\nabla }_B\zeta ^\delta )\zeta
^\alpha \zeta ^\beta \zeta ^\tau +\frac 12\nabla _\alpha T_{\left[ \delta
\beta \gamma \right] }(\widehat{\nabla }_A\zeta ^\beta )\epsilon ^{AB}(
\widehat{\nabla }_B\zeta ^\gamma )\zeta ^\alpha \zeta ^\delta \}. {%
\eqnum{3.8}}
\end{equation}
\narrowtext

Now we con\-struct the d-covariant la-background fun\-cti\-o\-nal  (we use
meth\-ods, in our case cor\-res\-pond\-ing\-ly adapted to the
N-con\-nect\-i\-on  structure, developed in [19,27,28]). The standard
quantization technique is based on the functional integral
\begin{equation}
Z\left[ J\right] =\exp \left( iW\left[ J\right] \right) =\int d[u]\exp
\{i\left( I+uJ\right) \},\stackrel{}{\eqnum{3.9}}
\end{equation}
with source $J^\alpha $ (we use condensed denotations and consider that
computations are made in the Euclidean space). The generation functional $%
\Gamma $ of one-particle irreducible (1PI) Green functions is defined as
$$
\Gamma \left[ \overline{u}\right] =W\left[ J\left( \overline{u}\right)
\right] -\overline{u}\cdot J\left[ \overline{u}\right] ,
$$
where $\overline{u}=\frac{\Delta W}{\delta J}$ is the mean field. For
explicit perturbative calculations it is useful to connect the source only
with the covariant quantum d-field $\zeta $ and to use instead of (3.9) the
new functional
\begin{equation}
\exp \left( iW\left[ \overline{u},J\right] \right) =\int \left[ d\zeta
\right] \exp \{i\left( I\left[ \overline{u}+v\left( \zeta \right) \right]
+J\cdot \zeta \right) \}.\stackrel{}{\eqnum{3.10}}
\end{equation}
It is clear that Feynman diagrams obtained from this functional are
d-covariant.

Defining the mean d-field $\overline{\zeta }\left( u\right) =\frac{\Delta W}{%
\delta J\left( u\right) }$ and introducing the auxiliary d-field $\zeta
^{\prime }=\zeta -\overline{\zeta }$ we obtain from (3.9) a double expansion
on both classical and quantum la-backgrounds:%
$$
\exp \left( i\overline{\Gamma [}\overline{u},\overline{\zeta ]}\right) =
$$
\begin{equation}
\int \left[ d\zeta ^{\prime }\right] \exp \{i\left( I\left[ \overline{u}%
+v\left( \zeta ^{\prime }+\overline{\zeta }\right) \right] -\zeta ^{\prime }
\frac{\Delta \Gamma }{\delta \overline{\zeta }}\right) \}.\stackrel{}{%
\eqnum{3.11}}
\end{equation}

The manner of fixing the measure in the functional (3.10) (and as a
consequence in (3.11) ) is obvious :
\begin{equation}
[d\zeta ]=\prod_u\sqrt{\mid G\left( u\right) \mid }\prod_{\alpha
=0}^{n+m-1}d\zeta ^\alpha \left( u\right) .\stackrel{}{\eqnum{3.12}}
\end{equation}
Using vielbein fields (3.6) we can rewrite the measure (3.12) in the form
$$
\left[ d\zeta \right] =\prod_u\prod_{\alpha =0}^{n+m-1}d\zeta ^{\underline{%
\alpha }}\left( u\right) .
$$
The structure of renormalization of $\sigma $-models of type (3.10) (or
(3.11)) is analyzed, for instance, in [19,27,28]. For la-spaces we must take
into account the N-connection structure.

\section{REGULARIZATION AND $\beta $-FUNCTIONS OF THE LOCALLY ANISOTROPIC $%
\sigma $-MODEL}

The aim of this section is to study the problem of regularization and
quantum ambiguities in $\beta $-functions of the renormalization group and
to present the results on one- and two-loop calculus for the la-$\sigma $%
-model (LAS-model).

\subsection{Regularization and Renormalization Group $\beta $-functions}

Because our $\sigma $-model is a two-dimensional and massless locally
anisotropic theory we have to consider both types of infrared and
ultraviolet regularizations (in brief, IR- and UV-regularization). In order
to regularize IR-divergences and distinguish them from UV-divergences we can
use a standard mass term in the action (3.1) of the LAS-model%
$$
I_{\left( m\right) }=-\frac{\tilde m ^2}{2\lambda ^2}\int d^2zG_{\alpha
\beta }u^\alpha u^\beta .
$$
For regularization of UF-divergences it is convenient to use the dimensional
regularization. For instance, the regularized propagator of quantum d-fields
$\zeta $ looks like%
$$
<\zeta ^\alpha \left( u_1\right) \zeta ^\beta \left( u_2\right) >=\delta
^{\alpha \beta }G\left( u_1-u_2\right) =
$$
$$
i\lambda ^2\delta ^{\alpha \beta }\int \frac{d^qp}{\left( 2\pi \right) ^q}
\frac{\exp \left[ -ip\left( u_1-u_2\right) \right] }{\left( p^2-\tilde m ^2+
i0\right) },
$$
where $q=2-2\epsilon .$

The d-covariant dimensional regularization of UF-divergences is complicated
because of existence of the antisymmetric symbol $\epsilon ^{AB}.$ One
introduces [29,27] this general prescription:%
$$
\epsilon ^{LN}\eta _{NM}\epsilon ^{MR}=\psi \left( \epsilon \right) \eta
^{LR}
$$
and
$$
\epsilon ^{MN}\epsilon ^{RS}=\omega \left( \epsilon \right) \left[ \eta
^{MS}\eta ^{NR}-\eta ^{MR}\eta ^{NS}\right] ,
$$
where $\eta ^{MN}$ is the q-dimensional Minkowski metric, and $\psi \left(
\epsilon \right) $ and $\omega \left( \epsilon \right) $ are arbitrary
d-functions satisfying conditions $\psi \left( 0\right) =\omega \left(
0\right) =1$ and depending on the type of renormalization.

We use the standard dimensional regularization , with dimensionless scalar
d-field $u^\alpha \left( z\right) ,$ when expressions for unrenormalized $%
G_{\alpha \beta }^{\left( ur\right) }$ and $B_{\alpha \beta }^{(k,l)}$ have
a d-tensor character, i.e. they are polynoms on d-tensors of curvature and
torsion and theirs d-covariant derivations (for simplicity in this
subsection we consider $\lambda ^2=1;$ in general one-loop 1PI-diagrams must
be proportional to $\left( \lambda ^2\right) ^{l-1}$).

RG $\beta $-functions are defined by relations (for simplicity we shall omit
index R for renormalized values)
$$
\mu \frac d{d\mu }G_{\alpha \beta }=\beta _{\left( \alpha \beta \right)
}^G\left( G,B\right) ,\mu \frac d{d\mu }B_{\left[ \alpha \beta \right]
}=\beta _{\left[ \alpha \beta \right] }^B\left( G,B\right) ,
$$
$$
\beta _{\alpha \beta }=\beta _{\left( \alpha \beta \right) }^G+\beta
_{\left[ \alpha \beta \right] }^B.
$$

By using the scaling property of the one-loop counter-term under global
conformal transforms%
$$
G_{\alpha \beta }^G\rightarrow \Lambda ^{\left( l-1\right) }G_{\alpha \beta
}^{\left( k,l\right) },B_{\alpha \beta }^{\left( k,l\right) }\rightarrow
\Lambda ^{\left( l-1\right) }B_{\alpha \beta }^{\left( k,l\right) }
$$
we obtain%
$$
\beta _{\left( \alpha \beta \right) }^G=-\sum_{l=1}^{(1,l)}lG_{(\alpha \beta
)}^{(1,l)},\beta _{[\alpha \beta ]}^B=-\sum_{l=1}^\infty lB_{[\alpha \beta
]}^{(1,l)}
$$
in the leading order on $\epsilon $ (compare with the usual perturbative
calculus from [31]).

The d-covariant one-loop counter-term is taken as%
$$
\Delta I^{\left( l\right) }=\frac 12\int d^2zT_{\alpha \beta }^{\left(
l\right) }\left( \eta ^{AB}-\epsilon ^{AB}\right) \partial _Au^\alpha
\partial _Bu^\beta ,
$$
where
\begin{equation}
T_{\alpha \beta }^{\left( l\right) }=\sum_{k=1}^l\frac 1{\left( 2\epsilon
^k\right) }T_{\alpha \beta }^{\left( k,l\right) }\left( G,B\right) .%
\stackrel{}{\eqnum{4.1}}
\end{equation}
For instance, in the three-loop approximation we have
\begin{equation}
\beta _{\alpha \beta }=T_{\alpha \beta }^{(1,1)}+2T_{\alpha \beta
}^{(1,2)}+3T_{\alpha \beta }^{(1,3)}.\stackrel{}{\eqnum{4.2}}
\end{equation}

In the next subsection we shall also consider constraints on the structure
of $\beta $-functions connected with conditions of integrability (caused by
conformal invariance of the two-dimensional world-sheet).

\subsection{One-loop Divergences and RG-equations of the LAS-model}

We generalize the one-loop results [32] to the case of la-backgrounds. If in
locally isotropic models one considers an one-loop diagram, for the
LAS-model the distinguished by N-connection character of la-interactions
leads to the necessity to consider four one-loop diagrams (see Fig. 1).To
these diagrams one corresponds counter-terms:%
$$
I_1^{\left( c\right)
}=I_1^{(c,x^2)}+I_1^{(c,y^2)}+I_1^{(c,xy)}+I_1^{(c,yx)}=
$$
$$
-\frac 12I_1\int d^2z\widehat{r}_{ij}\left( \eta ^{AB}-\epsilon ^{AB}\right)
\partial _Ax^i\partial _Bx^j-
$$
$$
\frac 12I_1\int d^2z\widehat{r}_{ab}\left( \eta ^{AB}-\epsilon ^{AB}\right)
\partial _Ay^a\partial _By^b-
$$
$$
\frac 12I_1\int d^2z\widehat{r}_{ia}\left( \eta ^{AB}-\epsilon ^{AB}\right)
\partial _Ax^i\partial _By^a-
$$
$$
\frac 12I_1\int d^2z\widehat{r}_{ai}\left( \eta ^{AB}-\epsilon ^{AB}\right)
\partial _Ay^a\partial _Bx^i,
$$
where $I_1$ is the standard integral%
$$
I_1=\frac{G\left( 0\right) }{\lambda ^2}=i\int \frac{d^qp}{(2\pi )^2}\frac
1{p^2- \tilde m ^2}=\frac{\Gamma \left( \epsilon \right) }{4\pi ^{\frac
q2}(\tilde m ^2)^\epsilon }=
$$
$$
\frac 1{4\pi \epsilon }-\frac 1\pi \ln \tilde m +\mbox{finite counter terms}%
.
$$

There are one-loops on the base and fiber spaces or describing quantum
interactions between fiber and base components of d-fields. If the
la-background d-connection is of distinguished Levi-Civita type we obtain
only two one-loop diagrams (on the base and in the fiber) because in this
case the Ricci d-tensor is symmetric. It is clear that this four-multiplying
(doubling for the Levi-Civita d-connection) of the number of one-loop
diagrams is caused by the ''indirect'' interactions with the N-connection
field. Hereafter, for simplicity, we shall use a compactified
(non-distinguished on x- and y-components) form of writing out diagrams and
corresponding formulas and emphasize that really all expressions containing
components of d-torsion generate four irreducible, see (2.14), types of
diagrams (with respective interaction constants) and that all expressions
containing components of d-curvature give rise in a similar manner to six
irreducible, see formulas (2.17), types of diagrams. We shall take into
consideration these details in the subsection where we shall write the
two-loop effective action.

Subtracting in a trivial manner $I_1,$
$$
I_1+subtractions=\frac 1{4\pi \epsilon },
$$
we can write the one-loop $\beta $-function in the form:%
$$
\beta _{\alpha \beta }^{\left( 1\right) }=\frac 1{2\pi }\widehat{r}_{\alpha
\beta }
$$
$$
=\frac 1{2\pi }\left( r_{\alpha \beta }-G_{\alpha \tau }T^{\left[ \tau
\gamma \phi \right] }T_{\left[ \beta \gamma \phi \right] }+G^{\tau \mu
}\nabla _\mu T_{\left[ \alpha \beta \tau \right] }\right) .
$$

We also note that the mass term in the action generates the mass one-loop
counter-term
$$
\Delta I_1^{(m)}=\frac{\tilde m ^2}2I_1\int d^2z\{\frac 13r_{\alpha \beta
}u^\alpha u^\beta -u_\alpha \tau _{.\beta \gamma }^\alpha G^{\beta \gamma
}\}.
$$

The last two formulas can be used for a study of effective charges as in
[22] where some solutions of RG-equations are analyzed. We shall not
consider in this paper such methods connected with the theory of
differential equations.

\subsection{Two-loop $\beta $-functions for the LAS-model}

In order to obtain two-loops of the LAS-model we add to the list (3.9) the
expansion\widetext
$$
\Delta I_{1\mid 2}^{\left( c\right) }=-\frac 12I_1\int d^2z\{\widehat{r}%
_{\alpha \beta }\left( \eta ^{AB}-\epsilon ^{AB}\right) \left( \widehat{%
\nabla }_A\zeta ^\alpha \right) \left( \widehat{\nabla }_B\zeta ^\beta
\right) +\left( \nabla _\tau \widehat{r}_{\alpha \beta }+\widehat{r}_{\alpha
\gamma }G^{\gamma \delta }T_{\left[ \delta \beta \tau \right] }\right)
\times
$$
$$
\left( \eta ^{AB}-\epsilon ^{AB}\right) \zeta ^\tau (\widehat{\nabla }%
_A\zeta ^\alpha )\partial _Bu^\beta +\left( \nabla _\tau \widehat{r}_{\alpha
\beta }-T_{\left[ \alpha \tau \gamma \right] }\widehat{r}_{.\beta }^\gamma
\right) \left( \eta ^{AB}-\epsilon ^{AB}\right) \partial _Au^\alpha \left(
\widehat{\nabla }_B\zeta ^\beta \right) \zeta ^\tau +
$$
$$
\left( \eta ^{AB}-\epsilon ^{AB}\right) (\frac 12\nabla _\gamma \nabla _\tau
\widehat{r}_{\alpha \beta }+\frac 12\widehat{r}_{\epsilon \beta }r_{\gamma
.\tau \alpha }^{.\epsilon }+\frac 12\widehat{r}_{\alpha \epsilon }r_{\gamma
.\tau \beta }^{.\epsilon }+T_{\left[ \alpha \tau \epsilon \right] }\widehat{r%
}^{\epsilon \delta }T_{\left[ \delta \gamma \beta \right] }+
$$
$$
G^{\mu \nu }T_{\left[ \nu \beta \gamma \right] }\nabla _\tau \widehat{r}%
_{\alpha \mu }-G^{\mu \nu }T_{\left[ \alpha \gamma \nu \right] }\nabla _\tau
\widehat{r}_{\mu \beta })\partial _Au^\alpha \partial _Bu^\beta \zeta ^\tau
\zeta ^\gamma \}
$$
and the d-covariant part of the expansion for the one-loop mass counter-term%
$$
\Delta I_{1\mid 2}^{\left( m \right) }=\frac{\left( \tilde m \right) ^2}%
2I_1\int d^2z\frac 13r_{\alpha \beta }\zeta ^\alpha \zeta ^\beta .
$$

The non-distinguished diagrams defining two-loop divergences are illustrated
in Fig.2. We present the explicit form of corresponding counter-terms
computed by using, in our case adapted to la-backgrounds, methods developed
in [27,29]:

For counter-term of the diagram $(\alpha ) $ we obtain
$$
(\alpha )=-\frac 12\lambda ^2(I_i)^2\int d^2z\{(\frac 14\Delta r_{\delta
\varphi }-\frac 1{12}\nabla _\delta \nabla _\varphi \overleftarrow r +\frac
12r_{\delta \alpha }r_\varphi ^\alpha -\frac 16r_{.\delta ._\varphi
}^{\alpha .\beta }r_{\alpha \beta +}
$$
$$
+\frac 12r_{._\delta }^{\alpha .\beta \gamma }r_{\alpha \varphi \beta \gamma
}+\frac 12G_{\delta \tau }T^{\left[ \tau \alpha \beta \right] }\Delta
T_{\left[ \varphi \alpha \beta \right] }+\frac 12G_{\varphi \tau }r_\delta
^\alpha T_{\left[ \alpha \beta \gamma \right] }T^{\left[ \tau \beta \gamma
\right] }-
$$
$$
\frac 16G^{\beta \tau }T_{\left[ \delta \alpha \beta \right] }T_{\left[
\varphi \gamma \tau \right] }r^{\alpha \gamma }+G^{\gamma \tau }T_{[\delta
\alpha \tau ]}\nabla ^{(\alpha }\nabla ^{\beta )}T_{\left[ \varphi \beta
\gamma \right] }+\frac 34G^{\kappa \tau }r_{.\delta }^{\alpha .\beta \gamma
}T_{\left[ \beta \gamma \kappa \right] }T_{\left[ \alpha \varphi \tau
\right] }-
$$
$$
\frac 14r^{\kappa \alpha \beta \gamma }T_{\left[ \delta \beta \gamma \right]
}T_{\left[ \kappa \alpha \varphi \right] })\partial _Au^\delta \partial
^Au^\varphi +\frac 14[\nabla ^\beta \Delta T_{\left[ \delta \varphi \beta
\right] }-3r_{...\delta }^{\gamma \beta \alpha }\nabla _\alpha T_{\left[
\beta \gamma \varphi \right] }-3T_{\left[ \alpha \beta \delta \right]
}\nabla ^\gamma r_{.._{\gamma \varphi }}^{\beta \alpha }+
$$
$$
\frac 14r^{\alpha \gamma }\nabla _\alpha T_{\left[ \gamma \delta \varphi
\right] }+\frac 16T_{\left[ \delta \varphi \alpha \right] }\nabla ^\alpha
r-4G^{\gamma \tau }T_{\left[ \tau \beta \delta \right] }T_{\left[ \alpha
\kappa \varphi \right] }\nabla ^\beta (G_{\gamma \epsilon }T^{\left[ \alpha
\kappa \epsilon \right] })+
$$
$$
2G_{\delta \tau }G^{\beta \epsilon }\nabla _\alpha (G^{\alpha \nu }T_{\left[
\nu \beta \gamma \right] })T^{\left[ \gamma \kappa \tau \right] }T_{\left[
\epsilon \kappa \varphi \right] }]\epsilon ^{AB}\partial _Au^\delta \partial
_Bu^\varphi \}.
$$

In order to computer the counter-term for diagram $(\beta ) $ we use
integrals:%
$$
\lim \limits_{u\rightarrow v}i<\partial _A\zeta \left( u\right) \partial
^A\zeta \left( v\right) >=i\int \frac{d^qp}{(2\pi )^q}\frac{p^2}{p^2- \tilde
m ^2} = \tilde m ^2I_1
$$
(containing only a IR-divergence) and
$$
J\equiv i\int \frac{d^2p}{(2\pi )^2}\frac 1{(p^2-\tilde m ^2)^2}=-\frac
1{(2\pi )^2}\int d^2k_E\frac 1{(k_E^2+ \tilde m ^2)^2}.
$$
(being convergent). In result we can express%
$$
(\beta )=\frac 16\lambda ^2\left( I_1^2+2\tilde m I_1J\right) \int d^2z
\widehat{r}_{\beta \alpha \gamma \delta }r^{\beta \gamma }\left( \eta
^{AB}-\epsilon ^{AB}\right) \partial _Au^\alpha \partial _Bu^\beta .
$$
In our further considerations we shall use identities (we can verify them by
straightforward calculations):%
$$
\widehat{r}_{(\alpha .\beta )}^{.[\gamma .\delta ]}=-\nabla _{(\alpha
}(G_{\beta )\tau }T^{[\tau \gamma \delta ]}),\widehat{r}_{\left[ \beta
\alpha \gamma \delta \right] }=2G^{\kappa \tau }T_{\left[ \tau [\alpha \beta
\right] }T_{\left[ \gamma \delta ]\kappa \right] ,}
$$
in the last expression we have three type of antisymmetrizations on indices,
$\left[ \tau \alpha \beta \right] ,\left[ \gamma \delta \kappa \right] $ and
$\left[ \alpha \beta \gamma \delta \right] ,$
$$
\widehat{\nabla }_\delta T_{\left[ \alpha \beta \gamma \right] }\widehat{%
\nabla }_\varphi T^{\left[ \alpha \beta \gamma \right] }=\frac 9{16}\left(
\widehat{r}_{\left[ \beta \alpha \gamma \right] \delta }-\widehat{r}_{\delta
\left[ \alpha \beta \gamma \right] }\right) \left( \widehat{r}%
_{...........\varphi }^{\left[ \beta \alpha \gamma \right] }-\widehat{r}%
_{...........\varphi }^{.\left[ \alpha \beta \gamma \right] }\right) -
$$
\begin{equation}
-\frac 94\widehat{r}_{\left[ \alpha \beta \gamma \delta \right] }\widehat{r}%
_{.........\varphi ]}^{[\alpha \beta \gamma }+\frac 94\widehat{r}%
_{......[\delta }^{\alpha \beta \gamma }\widehat{r}_{\left[ \varphi ]\alpha
\beta \gamma \right] }+\frac 94\widehat{r}_{.[\delta }^{\alpha .\beta \gamma
}\widehat{r}_{\left[ \varphi ]\alpha \beta \gamma \right] }.\stackrel{}{%
\eqnum{4.3}}
\end{equation}

The momentum integral for the first of diagrams $(\gamma ) $
$$
\int \frac{d^qpd^qp^{\prime }}{(2\pi )^{2q}}\frac{p_Ap_B}{(p^2-\tilde m
^2)([k+q]^2-\tilde m ^2)([p+q]^2- \tilde m ^2)}
$$
diverges for a vanishing exterior momenta $k_\mu .$The explicit calculus of
the corresponding counter-term results in
$$
\gamma _1=-\frac{2\lambda ^2}{3q}I_1^2\int d^2z\{(r_{\alpha (\beta \gamma
)\delta }+G^{\varphi \tau }T_{[\tau \alpha (\beta ]}T_{[\gamma )\delta
\varphi ]})\left( r_{.\mu }^{\beta .\gamma \delta }-G_{\mu \tau }G_{\kappa
\epsilon }T^{[\tau \beta \kappa ]}T^{\left[ \gamma \delta \epsilon \right]
}\right) \partial ^Au^\alpha \partial _Au^\mu +
$$
$$
(\nabla _{(\beta }T_{[\delta )\alpha \gamma ]})\nabla ^\beta (G_{\mu \tau
}T^{[\tau \gamma \delta ]})\epsilon ^{LN}\eta _{NM}\epsilon ^{MR}\partial
_Lu^\alpha \partial _Ru^\mu -
$$
\begin{equation}
2(r_{\alpha (\beta \gamma )\delta }+G^{\varphi \tau }T_{[\alpha \tau (\beta
]}T_{[\gamma )\delta \varphi ]})\nabla ^\beta (G_{\mu \epsilon }T^{[\epsilon
\delta \gamma ]})\epsilon ^{MR}\partial _Mu^\alpha \partial _Ru^\mu \}.
\stackrel{}{\eqnum{4.4}}
\end{equation}
The counter-term of the sum of next two $(\gamma )$-diagrams is chosen to be
the la-extension of that introduced in [29,27]:%
$$
\gamma _2+\gamma _3=-\lambda ^2\omega \left( \epsilon \right) \frac{10-7q}{%
18q}I_1^2\int d^2z\{\widehat{\nabla }_AT_{[\alpha \beta \gamma ]}\widehat{%
\nabla }^AT^{[\alpha \beta \gamma ]}+
$$
\begin{equation}
6G^{\tau \epsilon }T_{[\delta \alpha \tau ]}T_{[\epsilon \beta \gamma ]}
\widehat{\nabla }_\varphi T^{[\alpha \beta \gamma ]}\left( \eta
^{AB}-\epsilon ^{AB}\right) \partial _Au^\delta \partial _Bu^\varphi \}.%
\stackrel{}{\eqnum{4.5}}
\end{equation}

In a similar manner we can computer the rest part of counter-terms:%
$$
\delta =\frac 12\lambda ^2(I_1^2+m^2I_1J)\int d^2z\widehat{r}_{\alpha (\beta
\gamma )\delta }\widehat{r}^{\beta \gamma }\left( \eta ^{AB}-\epsilon
^{AB}\right) \partial _Au^\alpha \partial _Bu^\delta ,
$$
$$
\epsilon =\frac 14\lambda ^2I_1^2\int d^2z\left( \eta ^{AB}-\epsilon
^{AB}\right) [\Delta \widehat{r}_{\delta \varphi }+r_\delta ^\alpha \widehat{%
r}_{\alpha \varphi }+r_\varphi ^\alpha \widehat{r}_{\delta \alpha }
$$
$$
-2(G^{\alpha \tau }T_{\left[ \delta \beta \alpha \right] }T_{\left[ \varphi
\gamma \tau \right] }\widehat{r}^{\beta \gamma }-G_{\varphi \tau }T^{\left[
\tau \alpha \beta \right] }\nabla _\alpha \widehat{r}_{\delta \beta
}+G_{\delta \tau }T^{[\tau \alpha \beta ]}\nabla _\alpha \widehat{r}_{\beta
\varphi }]\partial _Au^\delta \partial _Bu^\varphi ,
$$
$$
\iota =\frac 16\lambda ^2 \tilde m ^2I_1J\int d^2z\widehat{r}_{\beta \alpha
\gamma \delta }r^{\beta \gamma }\left( \eta ^{AB}-\epsilon ^{AB}\right)
\partial _Au^\alpha \partial _Bu^\delta ,
$$
$$
\eta =\frac 14\lambda ^2\omega \left( \epsilon \right) \left( I_1^2+2\tilde
m ^2I_1J\right) \int d^2z\widehat{r}_{.\alpha \gamma \delta }^\beta
T_{\left[ \beta \varphi \tau \right] }T^{\left[ \gamma \varphi \tau \right]
}\left( \eta ^{AB}-\epsilon ^{AB}\right) \partial _Au^\alpha \partial
_Bu^\delta .
$$

By using relations (4.3) we can represent terms (4.4) and (4.5) in the
canonical form (4.1) from which we find the contributions in the $\beta
_{\delta \varphi }$-function (4.2):%
$$
\gamma _1:-\frac 2{3(2\pi )^2}\widehat{r}_{\delta (\alpha \beta )\gamma }
\widehat{r}_{........\varphi }^{\gamma (\alpha \beta )}-\frac{(\omega _1-1)}{%
(2\pi )^2}\{\frac 43\widehat{r}_{[\gamma (\alpha \beta )\delta ]}\widehat{r}%
_{........\varphi ]}^{[\gamma (\alpha \beta )}+\widehat{r}_{[\alpha \beta
\gamma \delta ]}\widehat{r}_{.......\varphi ]}^{[\alpha \beta \gamma }\},
$$
\begin{equation}
\gamma _2+\gamma _3:\frac{(4\omega _1-5)}{9(2\pi )^2}\{\widehat{\nabla }%
_\delta T_{\left[ \alpha \beta \gamma \right] }\widehat{\nabla }_\varphi
T^{\left[ \alpha \beta \gamma \right] }+6G^{\tau \epsilon }T_{\left[ \delta
\alpha \tau \right] }T_{\left[ \epsilon \beta \gamma \right] }\widehat{%
\nabla }_\varphi T^{\left[ \alpha \beta \gamma \right] }- -\frac{(\omega
_1-1)}{(2\pi )^2}\widehat{r}_{\left[ \alpha \beta \gamma \delta \right] }
\widehat{r}_{.......\varphi ]}^{[\alpha \beta \gamma }\},\stackrel{}{%
\eqnum{4.8}}
\end{equation}
$$
\eta :\frac{\omega _1}{(2\pi )^2}\widehat{r}^\alpha ._{\delta \beta \varphi
}T_{\left[ \alpha \tau \epsilon \right] }T^{\left[ \beta \tau \epsilon
\right] }.
$$
\narrowtext
Finally, in this subsection we remark that two-loop $\beta $-function can
not be written only in terms of curvature $\widehat{r}_{\alpha \beta \gamma
\delta }$ and d-derivation $\widehat{\nabla }_\alpha $ (similarly as in the
locally isotropic case [29,27]).

\subsection{Low-Energy Effective Action for La-Strings}

The conditions of vanishing of $\beta $-functions describe the propagation
of string in the background of la-fields $G_{\alpha \beta }$ and $b_{\alpha
\beta }.$ (in this section we chose the canonic d-connection $^{\circ}{\Gamma%
}^\alpha _{\cdot \beta \gamma}$ on ${\cal E ).}$ The $\beta $-functions are
proportional to d-field equations obtained from the on-shell string
effective action
\begin{equation}
I_{eff}=\int du\sqrt{|\gamma |}L_{eff}\left( \gamma ,b\right) .\stackrel{}{%
\eqnum{4.9}}
\end{equation}
The adapted to N-connection variations of (4.8) with respect to $G^{\mu \nu
} $ and $b^{\mu \nu }$ can be written as
$$
\frac{\Delta I_{eff}}{\delta G^{\alpha \beta }}=W_{\alpha \beta }+\frac
12G_{\alpha \beta }(L_{eff}+{\mbox{complete derivation)}},
$$
$$
\frac{\Delta I_{eff}}{\delta b^{\alpha \beta }}=0.
$$

The invariance of action (4.8) with respect to N-adapted diffeomorfisms gives
rise to the identity%
$$
\nabla _\beta W^{\alpha \beta }-T^{\left[ \alpha \beta \gamma \right] }\frac{%
\Delta I_{eff}}{\delta b^{\beta \gamma }}=
$$
$$
-\frac 12\nabla ^\alpha (L_{eff}+\mbox{complete derivation)}
$$
(in the locally isotropic limit we obtain the well-known results from
[33,34]). This points to the possibility to write out the integrability
conditions as
\begin{equation}
\nabla ^\beta \beta _{(\alpha \beta )}-G_{\alpha \tau }T^{\left[ \tau \beta
\gamma \right] }\beta _{[\beta \gamma ]}=-\frac 12\nabla _\alpha L_{eff}.%
\stackrel{}{\eqnum{4.10}}
\end{equation}
For one-loop $\beta $-function, $\beta _{\alpha \beta }^{\left( 1\right)
}=\frac 1{2\pi }\widehat{r}_{\alpha \beta },$ we find from the last equations%
$$
\nabla ^\beta \beta _{(\delta \beta )}^{(1)}-G_{\delta \tau }T^{[\tau \beta
\gamma ]}\beta _{[\beta \gamma ]}^{(1)}=\frac 1{4\pi }\nabla _\delta \left(
\overleftarrow R + \frac 13T_{\left[ \alpha \beta \gamma \right] }T^{\left[
\alpha \beta \gamma \right] }\right)
$$
We can take into account two-loop $\beta $-functions by fixing an explicit
form of
$$
\omega \left( \epsilon \right) =1+2\omega _1\epsilon +4\omega _2\epsilon
^2+...
$$
when
$$
\omega _{HVB}\left( \epsilon \right) =\frac 1{(1-\epsilon )^2},\omega
_1^{HVB}=1,\omega _2^{HVB}=\frac 34
$$
(the t'Hooft-Veltman-Bos prescription [35]). Putting values (4.8) into
(4.10) we obtain the two-loop approximation for la-field equations\widetext
$$
\nabla ^\beta \beta _{(\delta \beta )}^{(2)}-G_{\delta \tau }T^{[\tau \beta
\gamma ]}\beta _{[\beta \gamma ]}^{(2)}=\frac 1{2(2\pi )^2}\nabla _\delta
[-\frac 18r_{\alpha \beta \gamma \delta }r^{\alpha \beta \gamma \delta
}+\frac 14r_{\alpha \beta \gamma \delta }G_{\tau \epsilon }T^{[\alpha \beta
\tau ]}T^{[\epsilon \gamma \delta ]}+
$$
$$
\frac 14G^{\beta \epsilon }G_{\alpha \kappa }T^{[\alpha \tau \sigma
]}T_{[\epsilon \tau \sigma ]}T^{[\kappa \mu \nu ]}T_{[\beta \mu \nu ]}-\frac
1{12}G^{\beta \epsilon }G_{\gamma \lambda }T_{[\alpha \beta \tau
]}T^{[\alpha \gamma \varphi ]}T_{[\epsilon \varphi \kappa ]}T^{[\lambda \tau
\kappa ]}],
$$
which can be obtained from effective action%
$$
I^{eff}\sim \int \delta ^{n+m}u\sqrt{|\gamma |}[-\overleftarrow r+\frac
13T_{[\alpha \beta \gamma ]}T^{[\alpha \beta \gamma ]}-\frac{\alpha ^{\prime
}}4(\frac 12r_{\alpha \beta \gamma \delta }r^{\alpha \beta \gamma \delta }-
$$
$$
G_{\tau \epsilon }r_{\alpha \beta \gamma \delta }T^{[\alpha \beta \tau
]}T^{[\epsilon \gamma \delta ]}-G^{\beta \epsilon }G_{\alpha \upsilon
}T^{[\alpha \gamma \kappa ]}T_{[\epsilon \gamma \kappa ]}T^{[\upsilon \sigma
\varsigma ]}T_{[\beta \sigma \varsigma ]}+
$$
\begin{equation}
\frac 13G^{\beta \epsilon }G_{\delta \kappa }T_{[\alpha \beta \gamma
]}T^{[\alpha \delta \varphi ]}T_{[\varphi \upsilon \epsilon ]}T^{[\gamma
\upsilon \kappa ]})]\stackrel{}{.}{\eqnum{4.11}}
\end{equation}

The action (4.11) (for 2$\pi \alpha ^{\prime }=1\,$ and in locally isotropic
limit) is in good concordance with the similar ones on usual closed strings
[36,27].

We note that the existence of an effective action is assured by the
Zamolodchikov c-theorem [37] which was generalized [38] for the case of
bosonic nonlinear $\sigma $-model with dilaton connection. In a similar
manner we can prove that such results hold good for la-backgrounds.

\section{Scattering of gravitons and duality of locally anisotropic
sigma-models}

The quantum theory of la-strings can be naturally considered by using the
formalism of functional integrals on ''hypersurfaces'' (see Polyakov's works
[39]). In this section we study the structure of scattering amplitudes of
la-gravitons. Questions on duality of la-string theories will be also
analyzed.

\subsection{The generation functional of la-string amplitudes for scattering
of la-gravitons}

We introduce the Green function of la-tachyons, the fundamental state of
la-string, as an integral (after Weeck rotation in the Euclidean space )%
$$
G_t(p_{1,...,}p_N)=\int [D\gamma _{AB}(\zeta )][Du^\alpha (z)]\exp (-\frac
1{4\pi \alpha ^{\prime }}\int d^2z\sqrt{|\gamma |}\gamma ^{AB}\partial
_Au^\alpha \partial _Bu^\alpha )
$$
\begin{equation}
\int \left[ \prod_AD^2z_A\right] \sqrt{|\gamma (z_B)|}\exp (ip_B^\alpha
u(z_B)),\stackrel{}{\eqnum{5.1}}
\end{equation}
where the integration measure on $\gamma _{AB}$ includes the standard ghost
Fadeev-Popov determinant corresponding to the fixation of the
reparametrization invariance and%
$$
\left[ \prod_AD^2z_A\right] =\prod_{A\neq M,N,K}d^2z_A|z_M-z_{N\,}|^2\left|
z_K-z_M\right| ^2\left| z_K-z_N\right| ^2.
$$
Because the string quantum field theory can be uncontradictory formulated
for spaces of dimension d=26 we consider that in formula (5.1) $\alpha $
takes values from 0 to 25. Formula (5.1) leads to dual amplitudes for
la-tachyon scatterings for $p^2=\frac 4{\alpha ^{\prime }}$ (see [40] for
details and references on usual locally isotropic tachyon scattering).

The generating functional of Green functions (5.1) in the coordinate
u-representation can be formally written as a hyper surface mean value
$$
\Gamma ^0\left[ \Phi \right] =<\exp \left( -\frac 1{2\pi \alpha ^{\prime
}}\int d^2z\sqrt{|\gamma |}\Phi \left[ u\left( z\right) \right] \right) >.
$$
In order to conserve the reparametrization invariance we define the
la-graviton source as
$$
\Gamma ^0\left[ G\right] =<\exp \left( -\frac 1{4\pi \alpha ^{\prime }}\int
d^2z\sqrt{|\gamma |}\gamma ^{AB}\partial _Au^\alpha \partial _Bu^\beta
G_{\alpha \beta }[u\left( z\right) ]\right) >
$$
from which we obtain the Green function of a number of $K$ elementary
perturbations of the closed la-string ($K$ la-gravitons)
$$
G_g(u_1,...,u_K)=%
$$
$$
<\frac 12\int [\prod_{[j]}D^2z_{[j]}]\sqrt{|\gamma (z)|} G^{AB\;}(z_{[j]})%
\partial _Au^\alpha (z_{[j]})\partial _Bu^\beta (z_{[j]\;})\chi _{\alpha
\beta }^{\left( j\right) }\delta ^{(d)}(u_{[j]}-u(z_{[j]}))>,
$$
where $\chi _{\alpha \beta }^{\left( j\right) }$ are polarization d-tensors
of exterior la-gravitons and $[j] =1,2,...K.$ Applying the Fourier transform
we obtain
$$
G_g(p_1,...,p_K)=\int [D\gamma _{AB}\left( z\right) ][Du^\alpha \left(
z\right) ]\exp \left( -\frac 1{2\pi \alpha ^{\prime }}\int d^2z\sqrt{|\gamma
|}\frac 12G^{AB}\partial _Au^\alpha \partial _Bu^\alpha \right)
$$
$$
\int \left[ \prod_{[j]}D^2z_{[j]}\right] \frac 12\sqrt{|\gamma (z_{[j])}|}%
\gamma ^{AB}(z_{[j]})\partial _Au^\beta (z_{[j]})\partial _Bu^\gamma
(z_{[j]})\chi _{\alpha \beta }^{(j)}\exp [ip_{[j]}^\delta u^\delta (z)].
$$
Integrating the last expression on $G_{\mu \nu }\,$ and $u^\alpha $ for
d=26, when there are not conformal anomalies, we have%
$$
G_g(p_1,...,p_K)=\int \left[ \prod_{[j]}D^2z_{[j]}\right] (\partial _A\frac
\partial {\partial p_{[j]}}\cdot \chi ^{(j)}.\partial ^A\frac \partial
{\partial p_{[j]}})\times
$$
\begin{equation}
\int [D\sigma (z)]\exp [-\pi \alpha ^{\prime
}\sum_{[i,j]}p_{[i]}p_{[j]}V(z_{[i]},z_{[j]},\sigma )],\stackrel{}{%
\eqnum{5.2}}
\end{equation}
where V is the Green function of the Laplacian for the conformally-flat
metric $G_{AB}=e^\sigma \delta _{AB}:$%
$$
\partial _A(\sqrt{|\gamma |}\gamma ^{AB}\partial _A)V=-\delta
^2(z_{[i]}-z_{[j]})
$$
which can be represented as
$$
V(z_{[i]},z_{[j]},\sigma )=-\frac 1{4\pi }\ln
|z_{[i]}-z_{[j]}|^2,z_{[i]}\neq z_{[j]},
$$
$$
V\left( z_{[k]},z_{[k]}^p,\sigma \right) =\frac 1{4\pi }(\sigma
(z_{[k]})-\ln (\frac 1\epsilon )),
$$
for $\epsilon \,$ being the cutting parameter. \ Putting the last expression
into (5.2) we compute the Green function of la-gravitons:%
$$
G_g(p_1,...p_K)=\int \prod_{[j]\neq
[p],[q],[s]}d^2z_{[j]}|z_{[q]}-z_{[p]}|^2|z_{[s]}-z_{[q]}|^2|z_{[s]}-z_{[p]}|^2(\partial _A\frac \partial {\partial p_{[j]}}\cdot \chi ^{(j)}\cdot \partial ^A\frac \partial {\partial p_{[j]}})
$$
$$
\int d\sigma (z)\prod_{[i]<[m]}|z_{[i]}-z_{[m]}|^{\alpha ^{\prime
}p_{[i]}p_{[j]}}\exp \left[ -\frac 14\alpha ^{\prime
}\sum_{[k]}p_{[k]}^2\sigma (z_{[k]})\right] \exp \left[ \frac{\alpha
^{\prime }}4\ln (\frac 1\epsilon )\sum_{[k]}p_{[k]}^2\right] ,
$$
where the definition of integration on $\sigma (z)$ is extended as%
$$
\int d\sigma (z)\equiv \lim \limits_{\sigma _d(z_{[j]})\rightarrow -\infty
}^{\sigma _s(z_{[j]})\rightarrow +\infty }\prod_{[j]}\int_{\sigma
_d(z_{[j]})}^{\sigma _s(z_{[j]})}d\sigma (z_{[j]}).
$$
So the scattering amplitude
$$
A_g(p_1,...,p_K)=\lim \limits_{p_{[j]}^2\rightarrow
0}\prod_{[j]}p_{[j]}^2G_g(p_{1,}...,p_K)
$$
is finite if%
$$
\lim \limits_{\sigma _d(z_{[j]})\to -\infty ,p_{[j]}^2\to 0}|p_{[j]}^2\sigma
(z_{[j]})|=const<\infty .
$$
$G_g(p_1,...,p_{K)}\,$ has poles on exterior momenta corresponding to
massless locally anisotropic modes of spin 2 (la-gravitons). The final
result for the scattering amplitute of la-gravitons is of the form
$$
A_g(p_1,...,p_K)\sim \int \prod_{[j]\neq
[p],[q],[s]}d^2z_{[j]}|z_{[p]}-z_{[q]}|^2|z_{[s]}-z_{[q]}|^2|z_{[s]}-z_{[p]}|^2
$$
$$
\partial _A\frac \partial {\partial p_{[j]}^\alpha }\chi _{\alpha \beta
}^{(j)}\partial ^A\frac \partial {\partial p_{[j]}^\beta
}\prod_{[m]<[n]}|z_{[m]}-z_{[n]}|^{\alpha ^{\prime }p_{[m]}\cdot p_{[n]}}.
$$
If instead of polarization d-tensor $\chi _{\alpha \beta }^{(j)}$ the
graviton polarization tensor $\chi _{ik}^{(j)}$ is taken we obtain the well
known results on scattering of gravitons in the framework of the first
quantization of the string theory [2,41].

\subsection{Duality of La-$\sigma $-models}

Two theories are dual if theirs non-equivalent second order actions can be
generated by the same first order action. The action principle assures the
equivalence of the classical dual theories. But, in general, the duality
transforms affects the quantum conformal properties [42]. In this subsection
we shall prove this for the la-$\sigma $-model (3.1) when metric $\gamma $
and the torsion potential b on la-background $\xi $ do not depend on
coordinate u$^0.$ If such conditions are satisfied we can write for (3.1)
the first order action
$$
I=\frac 1{4\pi \alpha ^{\prime }}\int d^2z\{\sum_{\alpha ,\beta =1}^{n+m-1}[
\sqrt{|\gamma |}\gamma ^{AB}(G_{00}V_AV_B+2G_{0\alpha }V_A\left( \partial
_Bu^\alpha \right) +G_{\alpha \beta }(\partial _Au^\alpha )\left( \partial
_Bu^\beta \right) )+
$$
\begin{equation}
\epsilon ^{AB}(b_{0\alpha }V_B(\partial _Au^\alpha )+b_{\alpha \beta
}(\partial _Au^\alpha )(\partial _Bu^\beta ))]+\epsilon ^{AB}\widehat{u}%
^0(\partial _AV_B)+\alpha ^{\prime }\sqrt{|\gamma |}R^{(2)}\Phi (u)\},%
\stackrel{}{\eqnum{5.3}}
\end{equation}
where string interaction constants from (3.1) and (5.3) are related as $%
\lambda ^2=2\pi \alpha ^{\prime }.$

This action will generate an action of type (3.1) if we shall exclude the
Lagrange multiplier $\widehat{u}^0$ The dual to (5.3) action can be
constructed by substituting \ V$_A$ expressed from the motion equations for
fields V$_A$ (also obtained from action (5.3)):%
$$
\widehat{I}=\frac 1{4\pi \alpha ^{\prime }}\int d^2z\{\sqrt{|\gamma |}\gamma
^{AB}\widehat{G}_{\alpha \beta }(\partial _A\widehat{u}^\alpha )(\partial _B
\widehat{u}^\beta )+\epsilon ^{AB}\widehat{b}_{\alpha \beta }(\partial _A
\widehat{u}^\alpha )(\partial _B\widehat{u}^\beta )+\alpha ^{\prime }\sqrt{%
|\gamma |}R^{(2)}\Phi (u)\},
$$
where the knew metric and torsion potential are introduced respectively as
\narrowtext
$$
\widehat{G}_{00}=\frac 1{G_{00}},\widehat{G}_{0\alpha }=\frac{b_{0\alpha }}{%
G_{00}},\widehat{G}_{\alpha \beta }=G_{\alpha \beta }-\frac{G_{0\alpha
}G_{0\beta }-b_{0\alpha }b_{0\beta }}{G_{00}}
$$
and
$$
\widehat{b}_{0\alpha }=-\widehat{b}_{\alpha 0}=\frac{G_{0\alpha }}{G_{00}},
\widehat{b}_{\alpha \beta }=b_{\alpha \beta }+\frac{G_{0\alpha }b_{0\beta
}-b_{0\alpha }G_{0\beta }}{G_{00}}
$$
(in the formulas for the new metric and torsion potential indices $\alpha $
and $\beta $ take values 1,2,...n+m-1).

If the model (3.1) satisfies the conditions of one-loop conformal invariance
(see details for locally isotropic backgrounds in [3]), one holds these
la-field equations%
$$
\frac 1{\alpha ^{\prime }}\frac{n+m-25}3+[4(\nabla \Phi )^2-4\nabla ^2\Phi
-r-\frac 13T_{[\alpha \beta \gamma ]}T^{[\alpha \beta \gamma ]}]=0,
$$
$$
\widehat{r}_{(\alpha \beta )}+2\nabla _{(\alpha }\nabla _{\beta )}\Phi =0,
$$
\begin{equation}
\widehat{r}_{[\alpha \beta ]}+2T_{[\alpha \beta \gamma ]}\nabla ^\gamma \Phi
=0.\stackrel{}{\eqnum{5.4}}
\end{equation}
By straightforward calculations we can show that the dual theory has the
same conformal properties and satisfies the conditions (5.4) if the dual
transform is completed by the shift of dilaton field%
$$
\widehat{\Phi }=\Phi -\frac 12\log G_{00}.
$$

The system of la-field equations (5.4), obtained as a low-energy limit of
the la-string theory, is similar to Einstein-Cartan equations (2.28) and
(2.29). We note that the explicit form of locally anisotropic
energy-momentum source in (5.4) is defined from well defined principles and
symmetries of string interactions and this form is not postulated, as in
usual  locally isotropic field models, from some general considerations in
order to satisfy the necessary conservation laws on la-space whose
formulation is very  sophisticated because of nonexistence  of global and
even local group of symmetries of such type of spaces. Here we also remark
that the LAS-model with dilaton field interactions  does not generate in the
low-energy limit the Einstein-Cartan la-theory because the first system of
equations from (5.4) represents some constraints (being a consequence of the
two-dimensional symmetry of the model)  on torsion and scalar curvature
which can not be interpreted as some algebraic relations of type (2.29)
between locally anisotropic spin-matter  source and torsion. As a matter of
principle we can generalize our constructions by introducing interactions
with gauge la-fields and considering a variant of  locally anisotropic
chiral $\sigma$-model [43] in order to get a system of equations quite
similar to (2.29). However, there are not exhaustive arguments for favoring
the Einstein-Cartan theory and we shall not try in this work to generate it
necessarily from la-strings.

\section{Summary and Conclusions}

Let us try to summarize our results, discuss their possible implications and
make the basic conclusions. Firstly, we have shown that the Einstein-Cartan
theory has a natural extension for a various class of la-spaces. Following
the R. Miron and M. Anastesiei approach [10] to the geometry of la-spaces it
becomes evident the possibility and manner of formulation of classical and
quantum field theories on such spaces. Here we note that in la-theories we
have an additional geometric structure, the N-connection. From physical
point of view it can be interpreted, for instance,  as a fundamental field
managing the dynamics of splitting of high-dimensional space-time into the
four-dimensional and compactified ones. We can also consider the
N-connection as a generalized type of gauge field which reflects  some
specifics of la-field interactions and possible intrinsic structure of
la-spaces. It was convenient to analyze the geometric structure of different
variants of la-spaces (for instance, Finsler, Lagrange and generalized
Lagrange spaces) in order to make obvious physical properties and compare
theirs perspectives in developing of new physical models.

According to modern-day views the theories of fundamental field interactions
should be a low energy limit of the string theory. One of the main results
of this work is the proof of the fact that in the framework of la-string
theory is contained a more general, locally anisotropic, gravitational
physics. To do this we have developed the locally anisotropic nonlinear
sigma model and studied it general properties. We shown that the condition
of self consistent propagation of string on la-background impose
corresponding constraints on the N-connection curvature, la-space torsion
and antisymmetric d-tensor. Our extension of background field method for
la-spaces has a distinguished by N-connection character and the main
advantage of this formalism is doubtlessly its universality for all types of
locally isotropic or anisotropic spaces.

The presented one- and two-loop calculus for the LAS-model and used in this
work d-covariant dimensional regularization are developed for la-background
spaces modelled as vector bundles provided with compatible N-connection,
d-connection and metric structures. In the locally isotropic limit we obtain
the corresponding formulas for the usual nonlinear sigma model.

Finally, it should be stressed that we firstly calculated the amplitudes for
scattering of la-gravitons and that the duality properties of the formulated
in this work LAS-model are similar to those of models considered for locally
isotropic strings

\acknowledgments
I would like to thank Sergiu V.\ Ostaf for useful comments and discussions.

\widetext
\begin{figure}[htbp]
\begin{picture}(420,150) \setlength{\unitlength}{1pt}
\thicklines
\put(105,35){\circle{30}}
\put(120,35){\line(1,0){60}}
\put(120,32){\line(1,0){60}}
\put(120,38){\line(1,0){60}}
\put(130,10){\makebox(40,20){$I_1^{(m,x^2)}$}}
\put(125,40){\makebox(50,20){$\hat r {\partial x}^2$}}

\put(265,35){\circle{30}}
\put(280,35){\line(1,0){60}}
\put(280,32){\line(1,0){60}}
\put(280,38){\line(1,0){60}}
\put(290,10){\makebox(40,20){$I_1^{(m,y^2)}$}}
\put(285,40){\makebox(50,20){$\hat r {\partial y}^2$}}

\put(105,105){\circle{30}}
\put(120,105){\line(1,0){60}}
\put(120,102){\line(1,0){60}}
\put(120,108){\line(1,0){60}}
\put(130,80){\makebox(40,20){$I_1^{(m,xy)}$}}
\put(125,110){\makebox(55,20){$^1{\hat r}(\partial x)(\partial y)$}}

\put(265,105){\circle{30}}
\put(280,105){\line(1,0){60}}
\put(280,102){\line(1,0){60}}
\put(280,108){\line(1,0){60}}
\put(290,80){\makebox(40,20){$I_1^{(m,yx)}$}}
\put(285,110){\makebox(50,20){$^1{\hat r}(\partial y)(\partial x)$}}

\end{picture}
\caption{The Feyn\-man di\-a\-grams for the one-loop $\beta$-functions
of the LAS-model}

\begin{picture}(420,520) \setlength{\unitlength}{1pt}
\thicklines

\put(91,55){\circle{30}}
\put(67,55){\circle{16}}
\put(67,47){\line(0,1){16}}
\put(19,55){\line(1,0){40}}
\put(19,52){\line(1,0){40}}
\put(19,58){\line(1,0){40}}
\put(107,52){\line(1,0){60}}
\put(107,55){\line(1,0){60}}
\put(107,58){\line(1,0){60}}
\put(20,60){\makebox(40,20){$m^2 \hat r$}}
\put(111,60){\makebox(60,20){$\hat r (\eta - \varepsilon ) {\partial u}^2$}}
\put(50,00){\makebox(70,20){Diagram $(\iota )$}}

\put(130,150){\line(1,0){40}}
\put(130,147){\line(1,0){40}}
\put(130,153){\line(1,0){40}}
\put(115,150){\circle{30}}
\put(92,150){\circle{16}}
\put(14,150){\line(1,0){70}}
\put(14,147){\line(1,0){70}}
\put(14,153){\line(1,0){70}}
\put(86,144){\line(1,1){12}}
\put(86,155){\line(1,-1){12}}
\put(50,110){\makebox(70,20){Diagram $(\varepsilon )$}}
\put(14,154){\makebox(70,20){$\hat r (\eta -\varepsilon )^2(\partial u)^2$}}
\put(131,154){\makebox(40,20){$\hat r(\eta -\varepsilon )$}}

\put(415,150){\circle{30}}
\put(391,150){\circle{16}}
\put(202,150){\line(1,0){180}}
\put(202,147){\line(1,0){180}}
\put(202,153){\line(1,0){180}}
\put(385,144){\line(1,1){12}}
\put(385,155){\line(1,-1){12}}
\put(290,110){\makebox(70,15){Diagram $(\delta )$}}
\put(205,155){\makebox(175,15){$(\nabla ^2\hat r +r\hat r + T^2\hat r +
T\nabla \hat r)(\eta - \varepsilon )(\partial u)^2$}}

\put(325,55){\circle{30}}
\put(270,55){\line(1,0){110}}
\put(340,52){\line(1,0){40}}
\put(340,58){\line(1,0){40}}
\put(270,58){\line(1,0){40}}
\put(270,52){\line(1,0){40}}
\put(325,25){\line(0,1){15}}
\put(328,25){\line(0,1){15}}
\put(322,25){\line(0,1){15}}
\put(270,60){\makebox(40,20){$\varepsilon T$}}
\put(340,60){\makebox(50,20){$\varepsilon T$}}
\put(330,26){\makebox(70,15){$(\eta - \varepsilon ) (\partial u)^2$}}
\put(290,00){\makebox(70,15){Diagram $(\eta )$}}

\put(390,300){\circle{30}}
\put(405,300){\line(1,0){15}}
\put(405,297){\line(1,0){15}}
\put(405,303){\line(1,0){15}}
\put(360,300){\line(1,0){15}}
\put(360,297){\line(1,0){15}}
\put(360,303){\line(1,0){15}}
\put(390,270){\line(0,1){15}}
\put(387,270){\line(0,1){15}}
\put(393,270){\line(0,1){15}}
\put(370,250){\makebox(40,15){$\varepsilon T(\partial u)$}}
\put(406,305){\makebox(15,15){$\varepsilon T$}}
\put(360,305){\makebox(15,15){$\varepsilon T$}}
\put(343,296){\makebox(5,10){+}}

\put(300,300){\circle{30}}
\put(270,300){\line(1,0){60}}
\put(270,303){\line(1,0){15}}
\put(270,297){\line(1,0){15}}
\put(315,303){\line(1,0){15}}
\put(315,297){\line(1,0){15}}
\put(316,305){\makebox(15,15){$\varepsilon T$}}
\put(270,305){\makebox(15,15){$\varepsilon T$}}
\put(253,296){\makebox(5,10){+}}

\put(190,225){\makebox(70,15){Diagrams $(\gamma )$}}

\put(00,300){\line(1,0){240}}
\put(120,300){\circle{30}}
\put(00,303){\line(1,0){105}}
\put(00,297){\line(1,0){105}}
\put(135,303){\line(1,0){105}}
\put(135,297){\line(1,0){105}}
\put(00,305){\makebox(105,20){$(r+T^2+\varepsilon \nabla T)\partial u$}}
\put(136,305){\makebox(105,20){$(r+T^2+\varepsilon \nabla T)\partial u$}}

\put(00,470){\line(1,0){230}}
\put(230,485){\circle{28}}
\put(230,455){\circle{28}}
\put(00,467){\line(1,0){230}}
\put(00,473){\line(1,0){230}}
\put(05,450){\makebox(200,15){$(\nabla ^2r+r^2+T\nabla ^2T=rT^2)
(\partial u)^2+$}}
\put(05,430){\makebox(200,15){$\varepsilon (\nabla ^3T+r\nabla T+T^2\nabla T+
T\nabla r)(\partial u)^2$}}
\put(80,390){\makebox(70,15){Diagram $(\alpha )$}}

\put(345,485){\circle{28}}
\put(345,455){\circle{28}}
\put(310,470){\line(1,0){30}}
\put(310,473){\line(1,0){30}}
\put(310,467){\line(1,0){30}}
\put(300,390){\makebox(70,15){Diagram $(\beta )$}}
\put(360,485){\line(1,0){65}}
\put(360,482){\line(1,0){65}}
\put(360,488){\line(1,0){65}}
\put(315,475){\makebox(5,10){$r$}}
\put(361,465){\makebox(64,15){$\hat r(\eta -\varepsilon )(\partial u)^2$}}

\end{picture}
\caption{Two-loops diagrams for the LAS-model}
\end{figure}


\begin{references}
\bibitem{1} C.\ Lovelace,\ Phys.\ Lett.\ B {\bf 35,} 75 (1984).
\bibitem{2} S.\ Fradkin and A.\ A.\ Tseytlin, Phys.\ Lett.\ B {\bf 158,} 316
(1985); Nucl. Phys. B {\bf 261,} 1 (1985).
\bibitem{3} C.\ G.\ Callan,\ D.\ Friedan, E.\ J.\ Martinec, and M.\ J.\ Perry,\
Nucl.\ Phys.\ B {\bf 262,} 593 (1985).
\bibitem{4} A.\ Sen,\ Phys.\ Rev.\ Lett.\ {\bf 55,} 1846 (1986).
\bibitem{5} S.\ P.\ de Alwis,\ Phys.\ Rev.\ D {\bf 34,} 3760 (1986).
\bibitem{6} A.\ M.\ Polyakov,\ Phys.\ Lett.\ B {\bf 103,} 107 (1981).
\bibitem{7} P.\ Finsler, {\it Uber und Fl\"{a}chen in allgemeinen R\"{a}umen.}
Dissertation, G\"{o}tingen, 1918 (Verlag Birkh\"{a}ser, Basel, 1951).
\bibitem{8} L.\ Berwalld,\ Math.\ Z.\ {\bf 25,} 40 (1926).
\bibitem{9} E.\ Cartan, {\it Les Espases de Finsler,} (Hermann, Paris, 1934).
\bibitem{10} R.\ Miron and M.\ Anastasiei,\ {\it The Geometry of Lagrange Spaces:\
 Theory and Applications} (Kluwer, Dordrecht, 1994);\ R.\ Miron and M.\
Anastasiei, {\it Vector Bundles. Largange Spaces. Applications to Relativity}
(Editura Academiei Rom\u{a}ne, Bucharest, 1987)\  (in Romanian).
\bibitem{11} H.\ Rund, {\it The Differential Geometry of Finsler Spaces,}
 (Springer--Verlag, Berlin, 1959);
 G.\ S.\ Asanov,\ {\it Finsler Geometry, Relativity and Gauge
Theories,} (D. Reidel, Dordrecht, 1985);
G.\ S.\ Asanov and S.\ F.\ Ponomarenko
 {\it Finsler Bundles on Space-Time. Associated Gauge Fields
and Connections,} ( \c{S}tiin\c{t}a, Chi\c{s}in\u{a}u, 1988) \ (in Russian);
R.\ Miron and T.\ Kawaguchi, Int.\ J.\ Theor.\ Phys.\ {\bf 30,} 1521 (1991);
A.\ A.\ Vlasov,\ {\it Statistical Distribution Functions} (Nauka, Moscow, 1966)
 (in Russian); R.\ Pimenov,\ Spaces of Kinematic Type (Nauka, Leningrad, 1968)
\ (in Russian); G.\ Yu.\ Bogoslovskyi, The Theory of Locally Anisotropic
Space-Time,\ (Moscow Univrsity Press, Moscow, 1982) \ (in Russian);
A.\ Bejancu,\ {\it Finsler Geometry and Applications} (Ellis Horwood Chichester,
England, 1990).
\bibitem{12} S.\ Vacaru, J.\ Math.\ Phys.\ {\bf 37,} 508 (1996).
 
\bibitem{13} S.\ Vacaru and Yu.\ Goncharenko, Int.\ J.\ Theor.\ Phys.\
{\bf 34,} 1955 (1995).
\bibitem{14} S.\ Vacaru, Nonlinear Connections in Superbundles and Locally
Anisotropic Supergravity (in preparation).
\bibitem{15} J.\ Kern,\ Arch.\ Math.\ {\bf 25,} 438 (1974).
\bibitem{16} W.\ Barthel,\ J.\ Reine.\ Angew.\ Math.\ {\bf 212,} 120 (1963).
\bibitem{17} M.\ Anastasiei, An.\ \c{S}t. Univ. Ia\c{s}i,\ s I a {\bf 32,} 17
 (1986).
\bibitem{18} A.\ A.\ Tseytlin,\ Int.\ J.\ Mod.\ Phys.\ A {\bf 4,} 1257 (1989).
\bibitem{19} C.\ M.\ Hull,\ {\it Lectures on  Non-linear Sigma Models and
Strings} (DAMPT, Cambridge, England, 1987).
\bibitem{20} E.\ Witten,\ Commun.\ Math.\ Phys.\ {\bf 92,} 455 (1984).
\bibitem{21} J.\ Wess and B.\ Zumino,\ Phys.\ Lett.\ {\bf 37,} 95 (1971).
\bibitem{22} E.\ Braaten,\ T.\ L.\ Curtright,\ C.\ K.,
and Zachos,\ Nucl.\ Phys.\ B  {\bf 260,} 630 (1985).
\bibitem{23} A.\ A.\ Belavin,\ A.\ M.\ Polyakov, and A.\ B.\
 Zamolodchikov, Nucl.\ Phys.\ B {\bf 241,} 333 (1984).
\bibitem{24} S.\ V.\ Ketov, {\it Introduction into Quantum Theory of Strings
and Superstrings} (Nauka, Novosibirsk, 1990)\ (in Russian).
\bibitem{25} B.\ S.\ DeWitt, {\it Dynamical Theory of Groups and Fields}
(Gordon and Breach, London, 1965).
\bibitem{26} L.\ Alvarez-Gaume,\ D.\ Z.\ Freedman,\ and S.\ Mukhi,\
 Ann.\ Phys.\ {\bf 134,} 85 (1981).
\bibitem{27} S.\ V.\ Ketov, {\it Nonlinear Sigma-Models in Quantum Field Theory
and String Theory} (Nauka, Novosibirsk, 1992)\ (in Russian).
\bibitem{28} R.\ S.\ Howe,\ and K.\ S.\ Stelle, The Background Field Method and
the Nonlinear $\sigma$-model, Preprint IAS, 1986 (unpublished).
\bibitem{29} S.\ V.\ Ketov, Nucl.\ Phys.\ {\bf 294,} 813 (1987).
\bibitem{30} B.\ E.\ Fridling and E.\ A.\ M.\ Van de Ven,\ Nucl.\ Phys.\ B
{\bf 268,} 719 (1986).
\bibitem{31} G.\ t'Hooft,\ Nucl.\ Phys.\ B {\bf 61,} 455 (1973).
\bibitem{32} T.\ L.\ Curtright and C.\ K.\ Zachos,\ Phys.\ Rev.\ Lett.\
{\bf 53,} 1799 (1984).
\bibitem{33} D.\ Zanon,\ Phys.\ Lett.\ B {\bf 191,} 363 (1987).
\bibitem{34} C.\ G.\ Callan,\ I.\ R.\ Klebanov,\ and M.\ J.\ Perry,\
 Nucl.\ Phys.\ B {\bf 278,} 78 (1986).
\bibitem{35} G.\ t'Hooft and M.\ Veltman,\ Nucl.\ Phys.\ B {\bf 44,} 189 (1971);
 M.\ Bos,\ Ann.\ Phys.\ (USA) {\bf 181,} 177 (1988).
\bibitem{36} D.\ J.\ Gross and J.\ H.\ Sloan, Nucl.\ Phys.\ B {\bf 291,} 41
(1987).
\bibitem{37} A.\ B.\ Zamolodchikov, Pis'ma J.\ T.\ P.\ {\bf 43,} 565 (1986).
\bibitem{38} A.\ A.\ Tseytlin, Phys.\ Lett.\ B {\bf 194,} 63 (1987).
\bibitem{39} A.\ M.\ Polyakov, Phys.\ Lett.\ B {\bf 103,} 207 and 211 (1981);
{\it Gauge Fields and Strings,} Contemporary Concepts in Physics (Harwood
Acad.\ Publ.,\ Chur e.\ a.,\  Switzerland, 1987).
\bibitem{40} {\it Superstrings: the first 15 years of superstring theory:
reprints \& commentary} by J.\ H.\ Schwarz (World Sci.,\ Singapore e.\ a.\ 1985).
\bibitem{41}S.\ V. Ketov, Izvestya Vuzov. Fizica {\bf 31,} 17 (1988)\
(in Russian).
\bibitem{42}T.\ H.\ Busher, Phys.\ Lett.\ B {\bf 201,} 466 (1988).
\bibitem{43} I.\ Jack, Nucl.\ Phys.\ B {\bf 234,} 365 (1984).
\end{references}
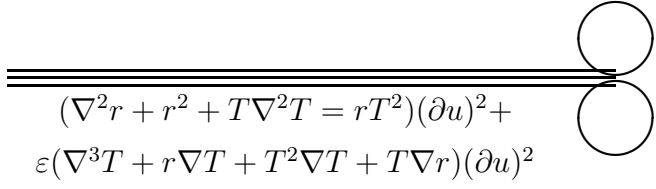
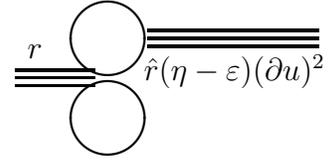
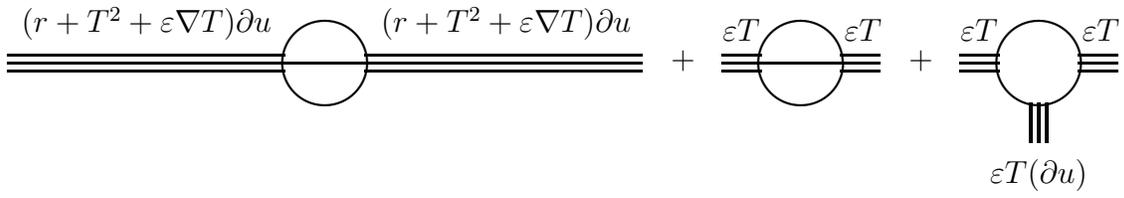
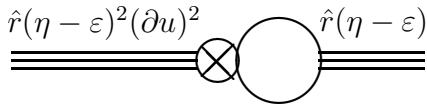
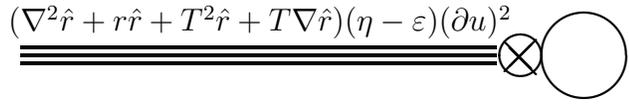
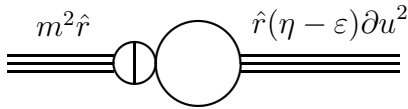
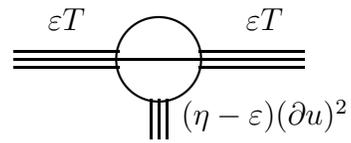
\end{document}